\documentclass[12pt,a4paper]{article}
\usepackage[dvips]{graphicx}
\usepackage{subfigure}
\usepackage{theorem}
\usepackage{xspace}
\usepackage {graphicx}
\usepackage{mathrsfs}
\usepackage[latin1]{inputenc}
\usepackage{amsmath}
\usepackage{amsfonts}
\usepackage{amssymb}
\usepackage{kformat}

\theorembodyfont{\upshape}

\newtheorem{example}{Example}

\newcommand{\diag}{\textrm{diag}}
\newcommand{\ds}{\displaystyle}
\newcommand{\ba}{\begin{equation}}
\newcommand{\ea}{\end{equation}}
\newcommand{\bea}{\begin{eqnarray}}
\newcommand{\eea}{\end{eqnarray}}
\newcommand{\bse}{\begin{subequations}}
\newcommand{\ese}{\end{subequations}}

\usepackage{setspace}
\begin{document}
\title{\textbf{Gauge Theory for Finite-Dimensional Dynamical Systems}}
\author{Pini Gurfil\\{\it Faculty  of Aerospace Engineering}\\{\it Technion - Israel Institute of Technology}\\{\it Haifa 32000, Israel}\\pgurfil@technion.ac.il}

\date{}

\maketitle \abstract{Gauge theory is a well-established concept in
quantum physics, electrodynamics, and cosmology. This theory has
recently proliferated into new areas, such as mechanics and
astrodynamics. In this paper, we discuss a few applications of gauge
theory in finite-dimensional dynamical systems with implications to
numerical integration of differential equations. We distinguish
between rescriptive and descriptive gauge symmetry. Rescriptive
gauge symmetry is, in essence, re-scaling of the independent
variable, while descriptive gauge symmetry is a Yang-Mills-like
transformation of the velocity vector field, adapted to
finite-dimensional systems. We show that a simple gauge
transformation of multiple harmonic oscillators driven by chaotic
processes can render an apparently ``disordered" flow into a regular
dynamical process, and that there exists a remarkable connection
between gauge transformations and reduction theory of ordinary
differential equations. Throughout the discussion, we demonstrate
the main ideas by considering examples from diverse engineering and
scientific fields, including quantum mechanics, chemistry,
rigid-body dynamics and information theory.}

\newpage
\section{Introduction}

In modern physics, gauge theories are probably among the most
powerful methods for understanding interactions among fields. The
importance of gauge theories for physics stems from the tremendous
success of the mathematical formalism in providing a unified
framework to describe the quantum field theories of
electromagnetism, the weak force and the strong force. Modern
theories like string theory, as well as some formulations of general
relativity, are, in some sense, gauge theories. The Yang-Mills
theory, the standard approach to quantum field theory, is a
particular example of gauge theories with non-Abelian symmetry
groups. Gauge symmetries are the core mathematical mechanism of
gauge theory, reflecting a redundancy in a description of a system.

The earliest physical theory which had a gauge symmetry was
Maxwell's electrodynamics. However, the importance of this symmetry
remained unnoticed in the earliest formulations. After Einstein's
development of general relativity, Hermann Weyl, in an attempt to
unify general relativity and electromagnetism , conjectured that
``Eichinvarianz" or invariance under the change of scale (or
``gauge") might also be a local symmetry of the theory of general
relativity. Weyl coined the use of gauge symmetry in modern physics
\cite{weyl1,weyl2,weyl3}, saying that ``symmetry, as wide or narrow
as you may define its meaning, is one idea by which man through the
ages has tried to comprehend and create order, beauty, and
perfection". 

The question to be raised at this point is: Can the
gauge-theoretical approach be applied on finite-dimensional systems,
and to what degree of success? Keeping in mind that Lie symmetry has
been a major tool in the study of ordinary differential equations
(ODEs) \cite{leach}, the answer must be positive; however, we have
not established yet a clear-cut connection between Lie symmetry and
gauge symmetry. We plan do so in the sequel, and show that, indeed,
Lie point-symmetry is closely connected to gauge symmetry, albeit
this connection is not always straightforward. In order to
facilitate the establishment of such a connection, we use gauge
symmetry in a more general context, a context of a symmetry defined
by diffeomorphisms. This will ultimately allow us to combine various
manifestations of gauge symmetry in finite-dimensional dynamical
systems under a single mathematical realm.

The idea to apply gauge transformations on ODEs is not new; it was
suggested by Kunin \cite{kunin0}, and has been recently revived by
Efroimsky
\cite{efroim1,efroim3,efroim4,efroimsky,efroimsky2,gurfil}, who
developed a gauge-generalized astrodynamical theory for modeling the
effect of orbital perturbations using non-osculating orbital
elements. Eforimsky's gauge theory, however, does not deal with
scale transformations, while Kunin's gauge theory is mostly
concerned with local gauge groups and discrete symmetries. Thus far,
there has not been a unified gauge theory which is able to support
both Yang-Mills-like gauge theories and scaling theories in
finite-dimensional systems.

In the current paper, we shall attempt to fill this gap by
developing a gauge theory for finite-dimensional dynamical systems
through two gauge symmetry mechanisms; the first symmetry mechanism
will be called \emph{rescriptive gauge symmetry}, evoked by carrying
out a \emph{rescriptive gauge transformation}.

Rescriptive gauge symmetry succumbs to the fundamental notion of
gauge transformations, namely, a change of scale, and is also
intimately connected to \emph{bilinearity}. To show rescriptive
gauge symmetry, we shall carry out an infinitesimal transformation
of the independent variable -- which in the bulk of our subsequent
discussion will be time -- into a different scale. The manifestation
of gauge symmetry in this case will be reflected in the ability to
obtain equivalence between the direction fields of the original and
gauge-transformed systems. In many practical applications, this
implies that the system can be reduced to a form amenable for
quadrature (e.~g. linear ODEs). We shall formalize this observation
by establishing a mechanism for \emph{reduction} through rescriptive
gauge symmetry.

To illustrate the concept of rescriptive gauge symmetry, we will
present a myriad of physical examples taken from diverse scientific
and engineering fields, including rigid-body dynamics,
finite-dimensional quantum mechanical systems, chemistry, and
information theory.

An instrumental constituent of our new theory is the \emph{gauged
pendulum}. Generally speaking, a gauged pendulum is a physical
system with a quadratic integral of motion, whose behavior in the
time domain can be arbitrary, although its phase space structure
remains invariant under a change of scale. This implies that after a
suitable scale transformation, harmonic oscillations will emerge. We
show that many physical systems can be either re-formulated to match
the formalism of the gauged pendulum, or are natural gauge pendulums
per-se; a classical example for a natural gauged pendulum is the
Euler-Poinsot system, to be subsequently analyzed.

We ultimately utilize the notion of a gauged pendulum to question
some common engineering misconceptions of chaotic and stochastic
phenomena, and show that seemingly ``disordered" (deterministic) or
``random" (stochastic) behaviors can be ``ordered", or, put
differently, evoke simple patterns \cite{marsden0,marsden} using an
infinitesimal transformation of the time scale.  This brings into
play the notion of \emph{observation} and \emph{observables}; we
show that temporal observations may be misleading when used for
chaos detection.

The second symmetry mechanism, reminiscent of the gauge symmetry
arising in Maxwell's equations \cite{marsdenweinstein} and its
generalization into the Yang-Mills field theory, will be referred to
as \emph{descriptive gauge symmetry}. The concomitant gauge
transformation will be called a \emph{descriptive gauge
transformation}. Descriptive gauge symmetry naturally arises in
Newtonian mechanical systems, and can be thought of as an invariance
of some configuration space under a gauge transformation of the
covariant derivative. In fact, descriptive gauge symmetry may be
best understood by relating it to the method of
variations-of-parameters (VOP), which is an analytical formalism for
solving inhomogeneous (forced) differential equations.

Euler invented the VOP method \cite{euler1,euler2} for treating
highly nonlinear problems emerging in celestial mechanics. However,
it was Lagrange who employed this method for deriving his system of
equations describing the evolution of the orbital elements
\cite{lagrange1,lagrange2,lagrange3}, known as Lagrange's Planetary
Equations.
The relation between the VOP method and descriptive gauge symmetry
can be explained as follows. According to the VOP method, the
integration constants of the homogeneous solution of a given ODE are
endowed with a time variation due to the presence of an external
force. However, the transformation from the state variables of the
original problem's phase space to the new state variables defined as
the time-varying constants involves an inherent freedom, which, in
practical calculations, can be removed by means of a user-defined
constraint. The constraint may be essentially arbitrary insofar as
it does not come into contradiction with the equations of motion
written for the variable ``constants."
The internal freedom emerges under the following circumstances:
First, one should perturb some $n$-dimensional differential
equation, and solve it by the VOP method (i.e., using the
unperturbed generic solution ${\vec q}(t,x_1\ldots x_n)$ as an
ansatz, and making its constants $x_i$ time-varying); second, the
number of ``constants" promoted to variables must exceed $n$. Thus,
when the said equations are written as equations for the new state
variables $x_i$, the number of these variables will exceed that of
equations; hence the internal freedom. Mathematically, this freedom
is analogous to the gauge symmetry in electrodynamics, while the
removal of this freedom by imposing an arbitrary constraint is
analogous to fixing of a gauge in the Maxwell theory.

From a practical standpoint, we show that descriptive gauge symmetry
may be used to considerably mitigate the numerical truncation error
of numerical integrations, and even ``symplectify" non-symplectic
integrators. We also show how a given system of ODEs can undergo a
reduction under a descriptive gauge symmetry transformation.

\section{Preliminaries and Definitions} \label{sec:definitions}

 Consider a finite-dimensional dynamical system whose dynamics
are modeled using  first-order vector differential equations of the
form
\begin{equation}\label{sys1}
    \frac{d\vec x}{dt} = \vec f (\vec x )
\end{equation}
where $\vec x\in \mathbb R^n$, $\vec f:\mathbb R^n \rightarrow
\mathbb R^n$. Assume that this system is subjected to structural
modifications resulting from some re-formulation of the phase space,
perturbations, control inputs, exogenous disturbances or modeling
uncertainties. We shall generalize these modifications under a
single mathematical umbrella which we call a \emph{rescription}. A
rescription operator in the time domain, $\mathfrak F_t$, acts upon
the vector field $\vec f$ in the following manner:
\begin{equation}\label{resc1}
    \mathfrak F_t\circ \vec f = \vec g(\vec x,\vec u(\vec x,t) )=\frac{d\vec
    x}{dt}
\end{equation}
The vector field $\vec u:\mathbb R^n\times \mathbb R
\rightarrow\mathbb R^m,\,m\le n$ will be called a \emph{rescriptor}.
The rescriptor may be either \emph{static} or \emph{dynamic}. In the
former case, one may write $\vec u = K(\vec x,t)$, with $K:\mathbb
R^n\times\mathbb R \rightarrow \mathbb R^m$, while in the latter
case the rescriptor constitutes a dynamical system of the general
form
\begin{subequations}
\label{sys3}
\begin{eqnarray}
&& \frac{d\vec u}{dt} = \vec h(\vec x,{d\vec x/dt},\vec y, \vec u )  \\
&& \frac{d\vec y}{dt} = \vec {h}_1(\vec y,\vec u ).
\end{eqnarray}
\end{subequations}
where $\vec y\in \mathcal M\subset\mathbb R^q$, $\mathcal M$ being
some compact differentiable manifold, and $\vec h_1:\mathcal
M\times\mathbb R^m\rightarrow \mathbb R^q$ is a vector field. By
restricting the dynamic rescriptor from dependance on ${\vec u}$, we
can unify the dynamic and static rescription, viz. a static
rescription can be defined as a special case of a dynamic
rescription. 

In most cases, the rescription modifies - sometimes intentionally,
such as in the case of control inputs - the fundamental properties
of the original system. These ``fundamental properties" may be, for
instance, integrability, symmetry and structure/volume-preserving
measures. The new properties of the rescribed system can be
investigated both in the time domain and in the phase space.

However, in some cases the rescription is merely an illusion; that
is, the rescription does not change the phase space and the
fundamental properties of the original system, although it could
modify the flow $\varphi(x_i(t=t_0),t)$. A classical example is the
action of the rotation group $\mathcal G_0=SO(n)$ on Lagrangians of
the form $\mathcal L = k\dot{\vec x}^2$, which remain left-invariant
under the transformation $\vec x\mapsto T\vec x$, $T\in SO(n)$. We
shall exclude these trivial occurrences from our discussion, and
will explore a more general setting. In this general setting, the
system is invariant under the action of some (possibly time-varying)
finite-dimensional \emph{gauge group}, $\mathcal G$.

\section{Rescriptive Gauge Symmetry}

We ask whether the system can be ``de-rescribed" by finding new
independent variables, $\tau_j$, possibly different for each
rescriptor component $u_i$,  satisfying
\begin{equation}\label{taud}
    d\tau_j= G_i(\vec x,u_i(\vec x,t), d \vec x, dt)
\end{equation}
for which
\begin{equation}\label{resc3}
    \mathfrak F_{\tau}\circ \vec g = \vec f(\vec x)={\vec
    x}^{\prime},
\end{equation}
where the operator $()^{\prime}$ denotes differentiation of each
$x_i$ with respect to some $\tau_j$,
\begin{equation}\label{}
    \vec{x}^{\prime} = \frac{dx_i}{d\tau_j},\quad i=1,\ldots,n,\quad
    j\in[1,\ldots,n].
\end{equation}
If $\exists d\tau_j,\,j\in[1,\ldots,n]$ satisfying (\ref{taud}) such
that (\ref{resc3}) holds, then we shall say that system
(\ref{resc1}) exhibits \emph{full rescriptive gauge
symmetry}\footnote{ The definition of \emph{rescriptive} is given in
Webster's Revised Unabridged Dictionary (1913): ``Pertaining to, or
answering the purpose of, a rescript; hence, deciding; settling;
determining".} under the \emph{rescriptive gauge transformation}
(\ref{taud}). In this case $\vec u$ becomes either a static or a
dynamic \emph{rescriptive gauge function}.

A \emph{rescriptive gauge symmetry of order $p$} or simply
\emph{partial rescriptive gauge symmetry} comes about when the
rescriptive gauge transformation de-rescribes only $p$ state
variables, $p< n$, viz.
\begin{equation}\label{resc2}
    \mathfrak F_{\tau}\circ g_i=  f_i(\vec x)={
    x_i}^{\prime},\quad i\in \mathbb N^p.
\end{equation}
In this case, if $t\in \mathfrak t =[0,\,t_f), t_f\le\infty$ and
$\tau_j\in\mathbb R, \,j\in[1,\ldots,p]$, then
$\exists\tau_{j0},t_0, x_i(t_0), x_i(\tau_{j0})$ such that the flow
satisfies the \emph{gauge homeomorphism}
\begin{equation}\label{}
    \varphi(x_i(t_0),t) = \varphi(x_i(\tau_{j0}),\tau_j)
\end{equation}
for $\mathfrak t\cap\mathbb R$, where the flow is interpreted as the
one-parameter group of transformations
\begin{equation}\label{}
    G_t:x_i(t_0)\to x_i(t),\quad G_\tau:x_i(\tau_0)\to x_i(\tau).
\end{equation}
The notion of rescriptive gauge symmetry has far-reaching
applications in ``ordering" seemingly ``disordered" phenomena,
solving ordinary differential equations (ODEs) and improving
numerical integration thereof. We shall illustrate these ideas
 by discussing a few examples of practical interest. We
embark on our quest by presenting the notion of a \emph{gauged
pendulum}, dwelt upon in the following subsection.

\subsection{The Gauged Pendulum}
Finite-dimensional systems can often be modeled by Hamiltonian
vector fields induced by a nominal Hamiltonian, $\mathcal H$, and a
perturbing Hamiltonian $\Delta \mathcal H$. Moreover, in ubiquitous
fields of science and engineering, $\mathcal H$ is comprised of $n$
uncoupled harmonic oscillators \cite{arribaselipe}, namely
\begin{eqnarray}\label{hamil1}
    \mathcal H[\vec q(t),\vec p(t)] &=& \frac{1}{2}\left(\vec p^T \vec p+\vec q^T\Omega \vec q\right)=\frac{1}{2}\sum_{i=1}^n
    \left(p_i^2+\omega_i^2 q_i^2\right)\nonumber  \\
    &=&\frac{1}{2} \sum_{i=1}^n\mathcal H_i[q_i(t),p_i(t)]
\end{eqnarray}
where $\Omega=\diag(\omega_1^2,\ldots,\omega_n^2)$,
\begin{equation}\label{}
    \vec q = [q_1,\ldots,q_n]^T,\quad \vec p = [p_1,\ldots,p_n]^T,
\end{equation}
are the generalized coordinates and conjugate momenta, respectively,
so that $(\vec{q},\vec{p})\in T^*Q$, where $Q$ is the configuration
space, $T^*Q$ is the cotangent bundle of $Q$, and $\dim{T^*Q}=2n$ is
the dimension of the phase space.

Hamilton's equations for $i=1,\ldots,n$ are then
\begin{subequations}
 \label{osci}
\begin{eqnarray}
  \dot q_i &=& p_i\\
  \dot p_i &=& -\omega_i^2 q_i.
\end{eqnarray}
\end{subequations}
Carrying out the point transformation into action-angle variables,
given by
\begin{equation}\label{aa}
    q_i = \sqrt{\frac{\Phi_i}{\omega_i}}\sin{\phi_i},\quad p_i =
    \sqrt{\Phi_j\omega_j}\cos{\phi_i},
\end{equation}
simplifies the Hamiltonian (\ref{hamil1}) even further, into
\begin{equation}\label{hamil2}
    \mathcal H[\vec \omega(t),\vec \Phi(t)] =
    \frac{1}{2}\vec{\omega}^T\vec{\Phi}=\frac{1}{2}\sum_{i=1}^n{\omega_i}{\Phi_i},
\end{equation}
where
\begin{equation}\label{}
    \vec \omega = [\omega_1,\ldots,\omega_n]^T,\quad \vec \Phi =
    [\Phi_1,\ldots,\Phi_n]^T.
\end{equation}
We note that from the topological standpoint, in both (\ref{hamil1})
and (\ref{hamil2}) $\mathcal H\in S^n\subset \mathbb R^{n+1}$ is
always homeomorphic to an $n$-ellipsoid, and $\mathcal H_i\in S^1$
is an integral. Moreover, the transformation $(\vec q,\,\vec
p)\mapsto(\vec \phi,\,\vec \Phi)$ is a (universal) covering map of
$S^n$.

It is clear that the Hamiltonian (\ref{hamil1}) is left-invariant
under the action of the rotation group $\mathcal G_0=SO(n)$ on $\vec
q$ and $\vec p$ if $\omega_i=\omega_0$. However, we shall seek a
 broader invariance of $\mathcal H$ with respect to the gauge group, $\mathcal G$,
which does not necessarily adhere to the $SO(n)$ symmetry.

To that end, let us choose an arbitrary (not necessarily smooth)
scalar field $u_i(\vec{q},\,\vec{p}):T^*Q\rightarrow \mathbb R$ to
serve as our rescriptor, coupling the dynamics of the $n$ pendulums,
and re-write (\ref{osci}) into the \emph{strictly bilinear
form}\footnote{A strictly bilinear system with respect to $\vec x$
and $\vec u$ has the structure $\dot{\vec{x}} = M\vec x \vec
u,\,M\in\mathbb R^{n\times n}$. See \cite{bruni} for details.} in
$[q,p]$ and $u$:
\begin{subequations}
 \label{oscimod}
\begin{eqnarray}
  \dot q_i &=& p_i u_i(\vec{q},\,\vec{p}) \label{one} \\
  \dot p_i &=& -\omega_i^2 q_i u_i(\vec{q},\,\vec{p}). \label{two}
\end{eqnarray}
\end{subequations}
Obviously, a constant of motion for each of the pairs $(q_i,\,p_i)$
would be
\begin{equation}\label{}
   \mathcal C_i=\frac{1}{2}
    \left(p_i^2+\omega_i^2 q_i^2\right),\quad i=1,\ldots,n
\end{equation}
although $\mathcal C_i$ is no longer the Hamiltonian. Nevertheless,
system (\ref{oscimod}) remains integrable regardless of the
particular form of $u_i$, since there are $n$ integrals for $n$
degrees-of-freedom. This can readily observed by performing the
(affine in $dt$) rescriptive gauge transformation
\begin{equation}\label{gf}
    d\tau_i = u_i(\vec p,\vec q) dt,
\end{equation}
which, on one hand, extends (\ref{oscimod}) into the state-space
model
\begin{subequations}
 \label{oscimod2}
\begin{eqnarray}
  \dot q_i &=& p_i u_i(\vec p,\vec q) \label{one} \\
  \dot p_i &=& -\omega_i^2q_i u_i(\vec p,\vec q) \label{two}\\
  \dot \tau_i &=& u_i(\vec p,\vec q),
\end{eqnarray}
\end{subequations}
but, on the other hand, transforms (\ref{oscimod}) back into the
simple harmonic oscillator form in the independent variables
$\tau_i$, assuming the symplectic structure
\begin{subequations}
 \label{oscimod3}
\begin{eqnarray}
   q^{\prime}_i &=& p_i    \\
   p_i^{\prime} &=& -\omega_i^2 q_i.
\end{eqnarray}
\end{subequations}
Thus, $u_i$ is a rescriptive gauge function $\forall i$, and
$\mathcal C_i$ can be interpreted as the Hamiltonian again, that is,
\begin{equation}\label{originalham}
    \mathcal H[\vec q(\tau_i),\,\vec p(\tau_i)] = \frac{1}{2}\sum_{i=1}^n C_i = \frac{1}{2}\sum_{i=1}^n
    (p_i^2+\omega_i^2
    q_i^2).
\end{equation}
%
%
%
%
%
%
%
%
%
%
%
%
%
In this example the transformation $t\mapsto\tau_i,i=1,\ldots,n$ is
therefore a static rescriptive gauge transformation. This means that
$u_i$ may be used to control the flow of $p_i$ and $q_i$ in the time
domain, but the persistence of the integrability under the
transformation (\ref{gf}) forces the system to exhibit the same
behavior as the harmonic oscillator in the modified times $\tau_i$
for each degree-of-freedom.

In order to generalize this concept and illustrate how rescriptive
gauge functions emerge in common physical systems, we must allow
$u_i$ to be an output of a dynamical system, giving rise to dynamic
rescription, defined in \S\ref{sec:definitions}. In this case,
system (\ref{oscimod}), written for each degree-of-freedom,
$i=1,\ldots,n$, becomes
\begin{subequations}
 \label{oscimod5}
\begin{eqnarray}
  \dot q_i &=& p_i u_i(\vec p,\vec q)   \\
  \dot p_i &=& -\omega_i^2 q_i u_i(\vec p,\vec q) \\
  \dot u_i &=& h_i(\vec p,\vec q,u_i,\vec y) \\
  \dot {\vec y} &=& \vec h_1(\vec u,\vec y).
\end{eqnarray}
\end{subequations}
Carrying out the rescriptive gauge transformation (\ref{gf}) reveals
a partial rescriptive gauge symmetry:
\begin{subequations}
 \label{oscimod5}
\begin{eqnarray}
    q_i^{\prime} &=& p_i   \\
   p_i ^{\prime} &=& -\omega_i^2 q_i    \\
   u_i ^{\prime} &=& \frac{1}{u_i} h_i(\vec p,\vec q,u_i,\vec y)
\end{eqnarray}
\end{subequations}
Thus, independently of the particular characteristics of the dynamic
(or static) rescriptive gauge function, $u_i$, the system re-assumes
the harmonic oscillator structure for $(q_i,p_i)$. This situation
can therefore be viewed as a generalization of the pendulum model.
The persistence of the harmonic oscillations under the rescriptive
gauge transformation  gives rise to the concept of a \emph{gauged
pendulum}. The gauged pendulum is a dynamical system whose flow
becomes periodic under the rescriptive gauge transformation,
although the flow of the original system may exhibit arbitrary
behavior in the time domain. Such systems arise in ubiquitous fields
of science and engineering. For example, the following model arises
in the study of quantum mechanical phenomenon (assuming a zero
decoherence coefficient) \cite{khaneja}:
\begin{eqnarray}
  \dot r_1  &=& -u_1(r_1,r_2)u_2(r_1,r_2)r_2 \\
  \dot r_2  &=& u_1(r_1,r_2)u_2(r_1,r_2)r_1.
\end{eqnarray}
This is obviously a gauged pendulum with the static rescriptor $u =
u_1u_2$. we shall subsequently dwell upon additional physical
examples.

 An alternative formulation of systems exhibiting partial rescriptive gauge symmetry
with a dynamic rescriptive gauge function may written as
\begin{subequations}
 \label{oscimod51}
\begin{eqnarray}
 \dot q_i &=& p_i u_i(\vec p,\vec q)  \\
  \dot p_i &=& -\omega_i^2 q_i u_i(\vec p,\vec q) \\
\  \dot u_i &=& h_i(\vec p,\vec q,u_i)
\end{eqnarray}
\end{subequations}
which becomes
\begin{subequations}
 \label{oscimod6}
\begin{eqnarray}
   q_i^{\prime} &=& p_i    \\
   p_i^{\prime} &=& -\omega_i^2 q_i \\
   u_i^{\prime}&=& \frac{1}{u_i}h_i(\vec p,\vec q,u_i)
\end{eqnarray}
\end{subequations}
after de-rescription using our standard rescriptive gauge
transformation. Here the rescriptor $u_i$  still constitutes a
dynamic rescriptive gauge, albeit it is not an output of an
auxiliary dynamical system anymore. In fact, if we relieve $h_i$
from direct dependance upon $u_i$, viz. $u_i=h_i(\vec p,\vec q)$,
then we uncover additional integrals of the motion, $\mathcal
K_{i}$, defined by the quadrature
\begin{equation}\label{}
    \mathcal K_{i}=\frac{1}{2}u_i^2-\int h_i[\vec q(\tau_i),\vec p(\tau_i)]d\tau_i.
\end{equation}
These new constants posses a clear meaning, revealed by writing
\begin{subequations}
\begin{eqnarray}\label{}
    &&\dot u_i = -\frac{\partial \mathcal K_i}{\partial \tau_i}=h_i\\
    &&\dot \tau_i = \frac{\partial \mathcal K_i}{\partial u_i} = u_i.
\end{eqnarray}
\end{subequations}
Hence, $\tau_i, u_i$ can be interpreted as generalized coordinates
and conjugate momenta, respectively, evolving on a $2n$-dimensional
symplectic manifold. $\mathcal K_i$ is then a Hamiltonian, and the
dynamics of $(u_i,\tau_i)$ is integrable. This remarkable structure
implies that if the time derivative of the rescriptor does not
explicitly depend upon the rescriptor, then the rescriptive gauge
transformation may be viewed as the symplectomorphism
\begin{equation}\label{symplect}
    \left[%
\begin{array}{c}
  \dot q_i = \frac{\ds \partial \mathcal H(q_i(t),p_i(t))}{\ds \partial p_i
  }\\\\
  \dot p_i = -\frac{\ds\partial \mathcal H(q_i(t),p_i(t))}{\ds\partial q_i }\\
\end{array}%
\right]\mapsto \left[%
\begin{array}{c}
  q_i^{\prime}  = \frac{\ds\partial \mathcal H(q_i(\tau_i),p_i(\tau_i))}{\ds\partial p_i }
  \\\\
  p_i^{\prime}  = -\frac{\ds\partial \mathcal H(q_i(\tau_i),p_i(\tau_i))}{\ds\partial q_i  }
  \\\\
  \dot u_i = -\frac{\ds\partial \mathcal K_i}{\ds\partial \tau_i}
  \\\\
  \dot \tau_i = \frac{\ds\partial \mathcal K_i}{\ds\partial u_i}\\
\end{array}%
\right].
\end{equation}
The formulation in (\ref{symplect}) is general, and is not limited
to Hamiltonians of the form (\ref{originalham}); rather, if
$\mathcal H_i(p_i,q_i)=const.$ is a given Hamiltonian, then
Hamilton's equations
\begin{equation}\label{}
    \dot q_i = \frac{\partial \mathcal H_i(p_i,q_i)}{\partial
    p},\quad     \dot p_i = -\frac{\partial \mathcal H_i(p_i,q_i)}{\partial q}
\end{equation}
undergoing a rescription
\begin{equation}\label{}
    \dot q_i = \frac{\partial \mathcal H_i(p_i,q_i)}{\partial
    p}u_i(\vec{p},\vec{q}),\quad     \dot p_i = -\frac{\partial \mathcal H_i(p_i,q_i)}{\partial
    q}u_i(\vec{p},\vec{q})
\end{equation}
will still possess still $\mathcal H_i(p_i,q_i)=const.$ as an
integral, and can be de-rescribed using the rescriptive gauge
transformation $d\tau = u_i(\vec p,\vec q)dt$ into
\begin{equation}\label{}
     q_i'= \frac{\partial \mathcal H_i(p_i,q_i)}{\partial
    p},\quad     \dot p'= -\frac{\partial \mathcal H_i(p_i,q_i)}{\partial
    q}.
\end{equation}
Finally, a slightly different formulation of the gauged pendulum
with a dynamics rescriptive gauge symmetry, to be illustrated in
\S\ref{seceuler}, may be written as
\begin{subequations}
 \label{oscimod9}
\begin{eqnarray}
 \dot q_i &=& k_1 p_i u_i(\vec p,\vec q)  \\
  \dot p_i &=& k_2 q_i u_i(\vec p,\vec q) \\
\  \dot u_i &=& h_i(\vec p,\vec q,u_i)
\end{eqnarray}
\end{subequations}
which becomes
\begin{subequations}
 \label{oscimod10}
\begin{eqnarray}
   q_i^{\prime} &=& k_{1i} p_i   \label{cone} \\
   p_i^{\prime} &=& k_{2i} q_i \label{ctwo} \\
   u_i^{\prime}&=& \frac{1}{u_i}h_i(\vec p,\vec q,u_i)
\end{eqnarray}
\end{subequations}
after de-rescription using the rescriptive gauge transformation
$\tau_i = u_i dt$. Here we have
\begin{equation}\label{}
   \mathcal C_i=\frac{1}{2}
    \left(k_{1i} p_i^2 -  k_{2i} q_i^2\right),\quad i=1,\ldots,n
\end{equation}
as integrals. However, an important caveat is that
(\ref{cone})-(\ref{ctwo}) may be viewed as a gauged pendulum only if
$k_{1i}k_{2i}<0$. Otherwise, the de-rescription will yield
hyperbolic motion in the variable $\tau$.

\subsection{Newtonian Systems}
\label{sec:newtonrescriptive} We shall now show that Newtonian
systems can be re-written into the gauged pendulum formalism. To
that end, consider the system
\bse \label{newton} \bea
\dot q &=& vp\\
\dot p &=& -vq\\
\dot v &=& f(x,v). \eea\ese
Performing the ``action-angle" transformation (cf. Eq.~(\ref{aa}))
$p=\cos x ,\,q=\sin x$ and re-writing (\ref{newton}) yields
\begin{subequations}
\label{sysnewt1}
\begin{eqnarray}
  \dot x &=& v \\
  \dot v &=& f(x,v),
\end{eqnarray}
\end{subequations}
which is a state space representation of the Newtonian system
\begin{equation}\label{}
    \ddot x = f(x,\dot x).
\end{equation}
We immediately observe that our rescriptor is the velocity, $v$.
Hence, when dealing with problems in the Newtonian context, the
rescriptive gauge function is simply the \emph{gauge velocity}.

Our Newtonian system therefore possesses a trivial partial
rescriptive gauge symmetry, found by performing the rescriptive
gauge transformation $d\tau = dx = v dt$. In other words, in the
Newtonian case $\tau = x$, and the de-rescribed system becomes
\bse\bea
 q^{\prime} &=& p \\
 p^{\prime} &=& - q \\
 v^{\prime}  &=& f(x,v)/v,
\eea\ese
where $()^{\prime}$ denotes differentiation with respect to $x$. If
$f(x,v)=f(x)$, then
\begin{equation}\label{}
    \mathcal K = \frac{v^2}{2}-\int f(x)dx
\end{equation}
is an integral, $\mathcal K$ is the Hamiltonian for the original
system (\ref{sysnewt1}), and $\mathcal H = (q^2+p^2)/2$ is the
Hamiltonian of $(q,p)$. Thus, any Newtonian system can be written in
the gauged pendulum form by extending the phase space dimension by
one. Consequently, any system whose state-space model is similar to
(\ref{newton}) is a Newtonian system in disguise.

\subsection{Eulerian Systems}
\label{seceuler}
In a body-fixed frame, the attitude dynamics of a
rigid body are usually formulated by means of the
\emph{Euler-Poinsot} equations. In a free-spin case, these equations
look as
\begin{equation}
\label{EPfree}
     \mathbb I \dot {\vec\omega}+\vec\omega\times\mathbb I\vec\omega =
     0,
\end{equation}
$\mathbb I$ being the inertia tensor and $\vec\omega =
[\omega_1,\,\omega_2,\omega_3]^T \in \mathfrak S$ is the body
angular velocity vector, where $\mathfrak S$ is the foliation
$\{(I\omega_1,I\omega_2,I\omega_3)|I_1\omega_1^2+I_2\omega_2^2+I_3\omega_3^2=G^2\})$,
$G$ being the total angular momentum.

Assuming that the body axes coincide with the principal axes
 of inertia,
 \ba
 \mathbb I = \textrm{diag}(I_1,I_2,I_3),
 \label{I}
 \ea
the Euler-Poinsot equations are
\bse \label{EP}
\begin{eqnarray}
  \dot\omega_1 &=& \sigma_1\omega_2\omega_3 \\
  \dot\omega_2 &=& \sigma_2\omega_1\omega_3 \\
  \dot\omega_3 &=& \sigma_3\omega_1\omega_2
\end{eqnarray}
\ese
where
\begin{equation}\label{}
    \sigma_1 = \frac{I_2-I_3}{I_1},\quad \sigma_2 =
    \frac{I_3-I_1}{I_2},\quad \sigma_3 = \frac{I_1-I_2}{I_3}.
\end{equation}
We shall now show that the Euler-Poinsot equations are a classical
example of the gauged pendulum concept with a dynamic rescriptive
gauge, exhibiting partial rescriptive gauge symmetry of order 2. To
that end, define the rescriptive gauge transformation
\begin{equation}\label{dtau2}
    d\tau = \omega_3dt
\end{equation}
and re-write (\ref{EP}) into
\bse \label{EP2}
\begin{eqnarray}
  \omega_1^{\prime} &=& \sigma_1\omega_2 \label{EP2a} \\
  \omega_2^{\prime} &=& \sigma_2\omega_1  \label{EP2b}\\
  \omega_3^{\prime} &=& \frac{\sigma_3}{\omega_3}\omega_1\omega_2 \label{EP2c},
\end{eqnarray}
\ese
which adheres to the gauged pendulum model (\ref{oscimod10}). Thus,
in the modified scale $\tau$, $\omega_1$ and $\omega_2$ will exhibit
harmonic oscillations with frequency $\sqrt{ |\sigma_1\sigma_2 |}$
if $\sigma_1\sigma_2<0$, given by
\begin{eqnarray}
  \omega_2(\tau) &=& \frac{-\sigma_2\omega_{10}\sin(\omega_0\tau_0)+\omega_{20}\omega_0\cos(\omega_0\tau_0)}{\omega_0}\cos(\omega_0\tau)\nonumber \\
                  &+& \frac{\sigma_2\omega_{10}\cos(\omega_0\tau_0)+\omega_{20}\omega_0\sin(\omega_0\tau_0)}{\omega_0}\sin(\omega_0\tau) \\
  \omega_1(\tau) &=& \frac{\sigma_2\omega_{10}\cos(\omega_0\tau_0)+\omega_{20}\omega_0\sin(\omega_0\tau_0)}{\sigma_2}\cos(\omega_0\tau) \nonumber
  \\ &-& \frac{\omega_{20}\omega_0\cos(\omega_0\tau_0)-\sigma_2\omega_{10}\sin(\omega_0\tau_0)}{\sigma_2}\sin(\omega_0\tau)
\end{eqnarray}
where $\omega_0=\sqrt{|\sigma_1\sigma_2|}$,
$\omega_{10}=\omega_1(\tau_0)$, $\omega_2(\tau_0)=\omega_{20}$.

 The solution for the dynamic rescriptor $\omega_3$ can now be easily solved by quadrature.\footnote{Note that here the rescriptor has units of angular velocity, while in the Newtonian case it was the velocity. We shall re-iterate on this issue in the following sections.}
 Since
\begin{equation}\label{}
    C  = \frac{1}{2}\omega_3^2 - \sigma_3\int\omega_1\omega_2 d\tau
\end{equation}
is an integral,
\begin{eqnarray}
 \omega_3  &=&  \sqrt{2 C+A\sigma_3\cos^2(\omega_0 \tau)+ B \sigma_3 \sin(\omega_0 \tau)
 \cos(\omega_0 \tau)}
\end{eqnarray}
where
\begin{eqnarray}
  A &=& \frac{-2 \sigma_2^2 \cos^2(\omega_0 \tau_0) \omega_{10}^2-4 \cos(\omega_0 \tau_0) \omega_{10} \omega_{20} \sin(\omega_0 \tau_0) \omega_0 \sigma_2}{\omega_0^2 \sigma_2}\nonumber \\
  &+&\frac{-\omega_0^2 \omega_{20}^2+2 \omega_0^2 \omega_{20}^2 \cos^2(\omega_0 \tau_0) +\sigma_2^2 \omega_{10}^2 } {\omega_0^2 \sigma_2}  \\
  B &=& \frac{-2 \sigma_2^2 \cos(\omega_0 \tau_0) \omega_{10}^2 \sin(\omega_0 \tau_0)+4 \cos(\omega_0 \tau_0)^2 \omega_{10} \omega_{20} \omega_0 \sigma_2}{ \omega_0^2
  \sigma_2}\nonumber\\&+&\frac{-2 \omega_{20} \omega_0 \sigma_2 \omega_{10}+2 \omega_0^2 \omega_{20}^2 \sin(\omega_0 \tau_0) \cos(\omega_0 \tau_0)   }{ \omega_0^2
  \sigma_2}.
\end{eqnarray}
From (\ref{dtau2}), the new independent variable is
\begin{equation}\label{}
    \tau = \int{\omega_3}dt.
\end{equation}
To understand its physical meaning, we recall that if we let the hat
map $\hvec{\phantom{v}}:\Rs\to\so3$ denote the usual Lie algebra
isomorphism that identifies $(\so3,\lbrac{\,}{\,})$ with
$(\Rs,\times)$, then
 \ba
 \widehat{\vec\omega} = - \dot R R^T
 \label{omegas}
 \ea
where
 \begin{equation}
 \label{C}
  \widehat{\vec\omega} =\left[%
 \begin{array}{ccc}
  0 & -\omega_3 & \omega_2 \\
  \omega_3 & 0 & -\omega_1 \\
  -\omega_2 & \omega_1 & 0  \\
\end{array}%
\right]
\end{equation}
and $R\in SO(3)$ is a rotation matrix from inertial to body
coordinates. If we take the rotation sequence $3\to 1\to 3$,
$\phi\to\theta\to\psi$, evaluation of (\ref{omegas}) will entail the
well-known expressions for the components of the vector of the body
angular velocity $\vec\omega$ in terms of the Euler angles rates
$\dot\phi,\,\dot\theta,\,\dot\psi$:
\bse
\begin{eqnarray}
  \omega_1 &=& \dot{\phi}\sin \theta
  \sin \psi+\dot{\theta}\cos \psi
   \label{omega1}\\
  \omega_2 &=&
  \dot{\phi}\cos \psi\sin \theta-\dot{\theta}\sin \psi
   \label{omega2}\\
  \omega_3 &=& \dot{\psi}+\dot{\phi}\cos \theta
   \label{omega3}
\end{eqnarray}
\ese
%
Thus,
\begin{equation}\label{}
\tau = \int{\omega_3} = \psi +\int\dot\phi\cos\theta dt,
\end{equation}
so the $\omega_i-\tau$ dynamics may be viewed as a solution for the
\emph{phase space} of the Eulerian system.

The rescription in the Eulerian case possesses an interesting
symmetry. In the above discussion, we detected the dynamic
rescriptor $\omega_3$ for the pair $(\dot\omega_1,\,\dot\omega_2)$,
but there are two other possible rescriptions: $\omega_1$ for
$(\dot\omega_2,\,\dot\omega_3)$ and $\omega_2$ for
$(\dot\omega_1,\,\dot\omega_3)$.

Finally, we note that the rescriptive gauge transformation
\emph{linearized} the Euler-Poinsot equations; observe that
(\ref{EP2a})-(\ref{EP2b}) are linear, and (\ref{EP2c}) is a simple
linear quadrature thereof in the variable $z = \omega_3^2$. We shall
further dwell upon this finding in \S\ref{sec:berkovich}.

\subsection{Other Common Systems Exhibiting Rescriptive Gauge Symmetry}
\label{sec:other}
 The gauged pendulum is a particular case of
systems exhibiting rescriptive gauge symmetry. However, there are
systems that exhibit rescriptive gauge symmetry, which cannot be
rendered periodic after a rescriptive gauge transformation.
Generally speaking, such systems cannot be conveniently described
using the Hamiltonian formalism, although they do posses integrals.
Consider, for illustration, the dynamical equations of two chemical
reactants, $A$ and $B$, whose concentrations evolve according to the
bilinear rate law \cite{clary}
\bse\label{chem0}
\begin{eqnarray}
  \frac{ d[A]}{dt} &=& k_1[A][B] \\
  \frac{ d[B]}{dt}  &=& k_2[A][B].
\end{eqnarray}
\ese
These can be de-rescribed using e.~g. $d\tau = [A] dt$, yielding the
linear equations
\bse \label{chem1}
\begin{eqnarray}
  \frac{ d[A]}{d\tau} &=& k_1[B] \\
  \frac{ d[B]}{d\tau}  &=& k_2[B],
\end{eqnarray}
\ese
so that
\begin{equation}\label{}
    [B(\tau)] = [B(\tau_0)]{\rm e}^{k_2\tau},\quad  [A(\tau)] =
    [B(\tau_0)]\frac{k_1}{k_2}({\rm e}^{k_2\tau}-1)+[A(\tau_0)].
\end{equation}
An integral for system (\ref{chem1}) is $\mathcal C =
[A]-k_1/k_2[B]$ , albeit this is not the Hamiltonian. Consequently,
an additional class of systems exhibiting rescriptive gauge symmetry
may be written as
\begin{subequations}
 \label{hyper1}
\begin{eqnarray}
  \dot q_i &=& q_i u_i(\vec p,\vec q)   \\
  \dot p_i &=& q_i u_i(\vec p,\vec q) \\
  \dot u_i &=& h_i(\vec p,\vec q,u_i,\vec y) \\
  \dot {\vec y} &=& \vec h_1(\vec u,\vec y).
\end{eqnarray}
\end{subequations}

\subsection{The One-Parameter Lie Symmetry Group}

Thus far we have not explicitly spelled out a relationship between
the rescriptive gauge transformation and Lie point-symmetry
transformations. This is the purpose of the following discussion.

To keep things simple, assume a 1-DOF gauged pendulum model with a
static rescriptor, $u(p,\,q)$:
\begin{subequations}
 \label{gp2}
\begin{eqnarray}
  \dot q &=& p  u (  p,  q)  \\
  \dot p &=& - q  u (  p,  q).
\end{eqnarray}
\end{subequations}
This set of equations can be analyzed by means of one-parameter
groups based upon infinitesimal transformations. We demand the
equation to be invariant under infinitesimal changes of the
independent variable $t$, but \emph{without} a simultaneous
infinitesimal changes of the dependent variables. This leads to the
Lie point-symmetry transformation
\bse \label{lie1}
\begin{eqnarray}
  p &\to& p \\
  q &\to& q \\
  t &\to& \tau = t+\epsilon\zeta(p,q).
\end{eqnarray}
\ese
We now apply (\ref{lie1}) on (\ref{gp2}) by following these stages:
First, we write
\begin{equation}\label{dqdtau}
    \frac{dq}{d\tau} = \frac{dq}{dt+\epsilon \left(\frac{\partial
    \zeta}{\partial q} dq +\frac{\partial \zeta}{\partial
    p}dp\right)}+O(\epsilon^2)
\end{equation}
and
\begin{equation}\label{dpdtau}
    \frac{dp}{d\tau} = \frac{dp}{dt+\epsilon \left(\frac{\partial
    \zeta}{\partial q} dq +\frac{\partial \zeta}{\partial
    p}dp\right)}+O(\epsilon^2).
\end{equation}
Expanding (\ref{dqdtau}) and (\ref{dpdtau}) into a Taylor series
with $\epsilon$ as a first-order small parameter we get
\begin{equation}\label{}
    \frac{dq}{d\tau} = \frac{dq}{dt} - \epsilon \frac{dq}{dt}
    \left(\frac{\partial \zeta}{\partial
    q}\frac{dq}{dt}+\frac{\partial\zeta}{\partial
    p}\frac{dp}{dt}\right)+O(\epsilon^2),
\end{equation}
\begin{equation}\label{}
    \frac{dp}{d\tau} = \frac{dp}{dt} - \epsilon \frac{dp}{dt}
    \left(\frac{\partial \zeta}{\partial
    q}\frac{dq}{dt}+\frac{\partial\zeta}{\partial
    p}\frac{dp}{dt}\right)+O(\epsilon^2).
\end{equation}
These yield a partial differential equation (PDE) for $\zeta(p,q)$,
\begin{equation}\label{}
     u^2(p,q) \epsilon\left(p\frac{\partial\zeta(p,q)}{\partial q}-q\frac{\partial\zeta(p,q)}{\partial
    p}\right)-u(p,q)+1 = 0,
\end{equation}
the solution thereof is
\begin{equation}\label{}
    \zeta(p,q) = -\int^p{
    \frac{u(\eta,\sqrt{c-\eta^2})-1}{\epsilon \sqrt{c-\eta^2}
    u^2(\eta,\sqrt{c-\eta^2})}d\eta} + c_0 c
\end{equation}
where $ c = p^2+q^2 $ and $c_0$ is an integration constant.
 To relate (\ref{dqdtau}) and (\ref{dpdtau}) to the generators of
 the infinitesimal transformation, we write
 \begin{equation}\label{}
    \tau = t+\epsilon \zeta(p,q) + \ldots = t+ \epsilon  \textrm X t +\ldots
\end{equation}
where the operator $\textrm X$ is given by
\begin{equation}\label{sym1}
    \textrm X = \zeta(p,q)\frac{\partial}{\partial t}
\end{equation}
In addition, due to the fact that (\ref{gp2}) is autonomous, it will
also exhibit Lie point-symmetry with generator
\begin{equation}\label{sym2}
    \textrm {X}_1 = \frac{\partial}{\partial t}.
\end{equation}
Symmetries (\ref{sym1}) and (\ref{sym2}) form an Abelian Lie algebra
$\mathfrak X$ with the Lie bracket $ [\textrm {X},\textrm {X}_1] =
0$.

In essence, this symmetry implies that the \emph{direction field} is
\begin{equation}\label{df}
    \frac{dp}{dq} = -\frac{q}{p},
\end{equation}
and is therefore homogenous, that is, invariant under all dilations
$(p,q)\mapsto (e^\lambda p, e^\lambda q),\,\lambda \in \mathbb R$,
which holds true for any dynamic or static rescriptive gauge
$u(p,q)$. The connection to the rescriptive gauge symmetry can now
be easily obtained via Arnold's theorem \cite{arnold}, stating that
if a one-parameter group of symmetries of a direction field is
known, the equation $dp/dq=f(p,q)$ can be integrated explicitly.
This is obvious for the direction field (\ref{df}) of the gauged
pendulum.
%
%
%
%

\subsection{Reduction using Rescriptive Gauge Symmetry}
\label{sec:berkovich}

It is a well-known fact in dynamical system theory that under
certain conditions, systems that exhibit symmetry are also reducible
\cite{marsdenratiu}. We shall discuss reduction in the context of
rescriptive gauge theory by following a few fundamental steps;
ultimately, we will show that rescriptive gauge symmetry allows to
reduce classes of nonlinear system into linear ODEs, solved by
simple quadratures.

We begin our quest for the manifestation of reduction in the realm
of rescriptive gauges by asking how a rescriptor for a given ODE can
be found. We shall then show that the answer to this question is
related to a more profound problem - that of \emph{exact
linearization} of ODEs, or, as we shall call it for clarity -
\emph{global linearization}. We shall dwell upon the latter issue
shortly, and will first address the more basic query.

Finding a rescriptive gauge transformation for a given ODE is
important, since it may allow quadrature in the modified time scale
by reduction into linear forms. Consider, for illustration, the
1-DOF gauged pendulum model
\begin{subequations}
 \label{gp}
\begin{eqnarray}
  \dot q &=& p  u (  p,  q)  \\
  \dot p &=& - q  u (  p,  q),
\end{eqnarray}
\end{subequations}
which is readily transformed into the ODE
\begin{equation}\label{thethe}
    \ddot q -\frac{\partial u(p,q)}{\partial q}p\dot
    q+u(p,q)q\left[\frac{\partial u(p,q)}{\partial p}+u(p,q)
    \right].
\end{equation}
Thus, any ODE that is written in the form (\ref{thethe}) can be
transformed into the de-rescribed gauged pendulum
$q^{\prime\prime}+q=0$ using the rescriptive gauge transformation
$d\tau = u dt$. However, usually the rescriptor, $u$, cannot be
easily found. Consider, for instance, the nonlinear ODE
\begin{equation}\label{ode1}
    \ddot q - \dot{q}^2 \cot q+q \sin^2 q=0,
\end{equation}
for which the rescriptive gauge transformation
\begin{equation}\label{sinq}
    d\tau = \sin q dt,
\end{equation}
reveals that (\ref{ode1}) is no more than a harmonic oscillator in
disguise, viz. $ q^{\prime\prime} + q =0$. However, one cannot
determine that $u = \sin q$ by observation. This calls for a more
rigorous methodology for finding the rescriptor.

To that end, consider a second-order ODE of the form
\begin{equation}\label{ode2}
    \ddot q + f(q)\dot q^2+b_1 u(q)\dot q +\psi(q)=0.
\end{equation}
When can this ODE be transformed into the linear form
\begin{equation}\label{ode3}
    q^{\prime \prime}+b_1 q^{\prime}+b_0 q+c = 0
\end{equation}
by a rescriptive gauge transformation
\begin{equation}\label{}
    d\tau = u(q)dt
\end{equation}
\emph{only}?
The answer lies in the theory of \emph{exact linearization}
\cite{berko}, which seeks a transformation rendering a nonlinear ODE
amenable for quadrature. We shall prefer the term \emph{global
linearization}, emphasizing that this method is conceptually
different from the common point linearization. We shall ultimately
use global linearization theory to help us track down the rescriptor
of a given ODE.

The theory of global linearization suggests that ODEs of the form
(\ref{ode2}) can be  globally linearized by a transformation of the
from
\begin{equation}\label{tarns}
    z = \beta\int u\exp\left(\int f dq\right)dq,\quad d\tau = u(q)dt
\end{equation}
where $\beta=const$, if and only if (\ref{ode3}) can be written in
the form
\begin{eqnarray}
&&\ddot q+f(q){\dot q }^2+b_1 u \dot q +u\exp\left(-\int
f(q)dq\right)\nonumber \\ &\cdot& \left[b_0\int u\exp\left(\int
f(q)dq\right)dq+\frac{c}{\beta}\right]=0 \label{trans1},
\end{eqnarray}
This fundamental result can be adapted to the case in question. In
particular, since we are probing the case of rescriptive gauge
transformations, we must require that $z=q$, or, in other words,
that
\begin{equation}\label{assumptions}
    \beta=1,\quad u=u(q),\quad f=-\frac{1}{u}\frac{du}{dq}.
\end{equation}
In our discussion we allowed $u$ to be a function of both $q$ and
$p$, while (\ref{tarns}) permit a $u$ which is a function of $q$
only. Thus, we must take $u=u(q)$, as written in
(\ref{assumptions}). Relations (\ref{assumptions}) modify
(\ref{trans1}) into 
%
%
\begin{eqnarray}\label{}
&&\ddot q -\frac{1}{u}\frac{du}{dq}{\dot q }^2+b_1 u \dot q
+u\exp\left(-\int
f(q)dq\right)\left[b_0q+\frac{c}{\beta}\right] \\
&=& \ddot q -\frac{1}{u}\frac{du}{dq}{\dot q }^2+b_1 u \dot q
+u\exp\left(-\int
\frac{1}{u}\frac{du}{dq}dq\right)\left[b_0q+\frac{c}{\beta}\right] \\
&=& \ddot q -\frac{1}{u}\frac{du}{dq}{\dot q }^2+b_1 u \dot q
+u^2b_0q+\frac{c}{\beta}  = 0
\end{eqnarray}
Thus, we have proven that a second-order ODE can be transformed into
a linear ODE using a rescriptive gauge transformation (assuming that
the rescriptor is a function of the coordinate only) \emph{if and
only if} this ODE can be written as
\begin{equation}\label{easy0}
    \ddot q -\frac{1}{u(q)}\frac{du(q)}{dq}{\dot q }^2+b_1 u(q) \dot q
+u(q)^2b_0q+\frac{c}{\beta}  = 0.
\end{equation}
Eq.~(\ref{easy0}) immediately yields the rescriptor: It is the
square root of the coefficient of the coordinate, $q$, divided by
$\sqrt{b_0}$.

Returning to example (\ref{ode1}), we see that it succumbs to the
general from (\ref{easy0}) by substituting
\begin{equation}\label{}
    b_1 = 0,\quad b_0=1,\quad c = 0,
\end{equation}
which yields
\begin{equation}\label{}
    \ddot q -\frac{1}{u}\frac{du}{dq}{\dot q }^2
+u^2 q   = 0
\end{equation}
and it is immediately apparent that the rescriptor is $u=\sin q$,
which agrees with (\ref{sinq}). As a simple verification, we also
note that
\begin{equation}\label{}
    \frac{1}{u}\frac{du}{dq} = \cot q.
\end{equation}
%
%
%
Similarly, the chemical rate equations (\ref{chem0}), written as the
single ODE
\begin{equation}\label{}
    \frac{d^2[A]}{dt^2}-\frac{1}{[A]}\left(\frac{d[A]}{dt}\right)^2-k_2[A]\frac{d[A]}{dt}=0
\end{equation}
may be linearized using the transformation $d\tau=[A]dt$, as was
done in \S \ref{sec:other}.

 The above process can be repeated for higher-order ODEs
as well. The bottom line is that the theory of global linearization
is a convenient method for finding a rescriptor of a given ODE, or,
in other words, reduce it into a linear ODE using a rescriptive
gauge transformation.

To conclude this section, we shall show that there are well-known
ODEs that can be transformed into the reducible form (\ref{easy0})
using an additional auxiliary variable transformation. This
observation is inspired by \S\ref{seceuler}, where we have shown
that the Euler-Poinsot equations are transformed into a linear from
in the independent variable $\tau$ using the rescriptive gauge
transformation $d\tau = \omega_3 dt$ and the auxiliary
transformation $z=\omega_3^2$. For example, consider the ODE:
\begin{equation}\label{}
    \ddot q + q\dot q+kq^3 =0,\quad k = const.
\end{equation}
This ODE arises in a few practical problems \cite{lemmer}. To render
it globally linearizable using a rescriptive gauge transformation,
perform the auxiliary variable transformation $z=q^2$, so the
modified system reads
\begin{equation}\label{zeq}
    \ddot z -\frac{1}{2z}\dot z^2+\sqrt{z}\dot z +kz^2 =0.
\end{equation}
In this form, (\ref{zeq}) adheres to ansatz (\ref{easy0}), with the
rescriptor $u=\sqrt{z}=q$ and $k=b_0, \,b_1=1,\,c=0$. The
rescriptive gauge transformation $d\tau = \sqrt{z} dt$ transforms
(\ref{zeq}) into
\begin{equation}\label{}
     z''+ z' + 2kz = 0.
\end{equation}

%

%

\subsection{Illustrative Examples}
We shall now illustrate the rescriptive gauge transformation
formalism and the resulting gauged pendulum concept using a few
numerical examples.
\begin{example}[A damped pendulum is a gauged pendulum]\end{example}
Consider the model \cite{kunin1}:
\begin{subequations}
\label{damped}
\begin{eqnarray}
    \dot q &=& up \\
    \dot p &=& -uq \\
    \dot u &=& -\omega_0^2q-au.\label{31c}
\end{eqnarray}
\end{subequations}
By carrying out the transformation $p=\cos \phi ,\,q=\sin \phi$,
these equations are immediately recognized as a state-space model
for a damped nonlinear pendulum,
\begin{equation}\label{}
    \ddot{\phi}+\omega_0^2\sin{\phi}+a\dot\phi=0.
\end{equation}
%
%
%
%
%
System (\ref{damped}) complies with the gauged pendulum formalism
(\ref{oscimod51}); it can be therefore viewed as a \emph{rescribed
harmonic oscillator}, revealed by the rescriptive gauge
transformation $d\phi = u dt$, so that $\tau = \phi$:
\begin{equation}
\label{harmonic}
     q^{\prime} =  p, \quad      p^{\prime} = -q.
\end{equation}
Obviously, the rescriptor, or gauge velocity, is simply the angular
velocity, i.~e. $u=\dot\phi$. The scalar differential equation for
this dynamic rescriptive gauge function, Eq.~(\ref{31c}), assumes
the nonautonomous form
\begin{equation}\label{up}
         u^{\prime} =-\omega_0^2\sin(\phi)/u-a.
\end{equation}
For $a=0$, the rescriptive gauge function does not explicitly depend
upon the rescriptor itself, and (\ref{up}) is easily solved by
quadrature:
\begin{equation}\label{}
    u(\phi) =
    \sqrt{2\omega_0^2(\cos \phi -\cos \phi_0 )+u^2(\phi_0)}.
\end{equation}
It is interesting to note that under the rescriptive gauge symmetry,
the harmonic oscillator and the damped nonlinear pendulum are
represented by the \emph{same} mathematical formalism - although for
different independent variables - whereas the time flow of these
models is completely different. The harmonic oscillator, which is a
conservative system, does not have an attractor, since the motion is
periodic. The damped pendulum, on the other hand, is a dissipative
dynamical systems, in which volumes shrink exponentially, so its
attractor has $0$ volume in phase space. This alleged paradox stems
from the fact that the dissipative time flow of the damped pendulum
becomes periodic under a change of the independent variable. Thus,
an observer measuring the ``time", $\phi$, is bound to observe
periodic behavior, while an observer measuring the ``true" time,
$t$, will observe exponential decay.

\begin{center}
\begin{figure}[h]
\includegraphics[width=6in]{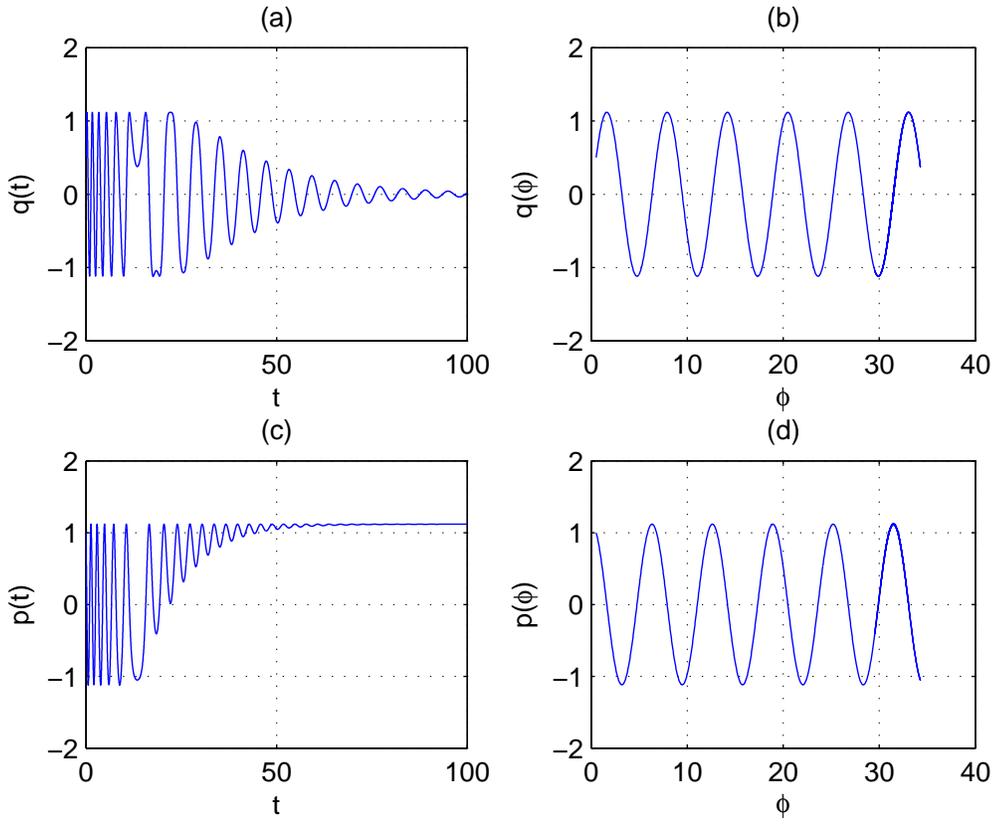}
\caption{An exponential decay of a damped nonlinear pendulum can be
transformed into harmonic oscillations by a rescriptive gauge
transformation.} \label{fig1}
\end{figure}
\end{center}

These observations are demonstrated and validated by means of a
numerical integration, comparing the flows of (\ref{damped}) and
(\ref{harmonic}). Figure \ref{fig1} compares between $q(t)$
(Fig.~\ref{fig1}a) and $q(\phi)$ (Fig.~\ref{fig1}b), and between
$p(t)$ (Fig.~\ref{fig1}c) and $p(\phi)$ (Fig.~\ref{fig1}d), for
$a=0.1,\, q_0 = 0.5,\,p_0 = 1,\,u_0 = 5,\,\phi_0=\sin^{-1}q_0 =
0.5236$.

\newpage
\begin{example}[A glimpse of order in the realm of chaos]\end{example}
Consider the dynamical system
\bse \label{lorentz}
\begin{eqnarray}
  \dot q &=& yp \label{lorentz1}\\
  \dot p &=& -yq \label{lorentz2}\\
  \dot x &=& \sigma (y-x) \label{l1}\\
  \dot y &=& (r-z)x-y \label{l2}\\
  \dot z &=& xy - bz \label{l3}
\end{eqnarray}
\ese
where $\sigma,\,r,\,b$ are constants. Eqs.~(\ref{l1})-(\ref{l3}) are
recognized as the Lorenz system, and the entire system
(\ref{lorentz}) complies with the gauged pendulum formalism of
Eqs.~(\ref{oscimod5}). It shall be thus referred to as the
\emph{Lorenz-fed gauged pendulum}.

\begin{center}
\begin{figure}[h]
\includegraphics[width=6in]{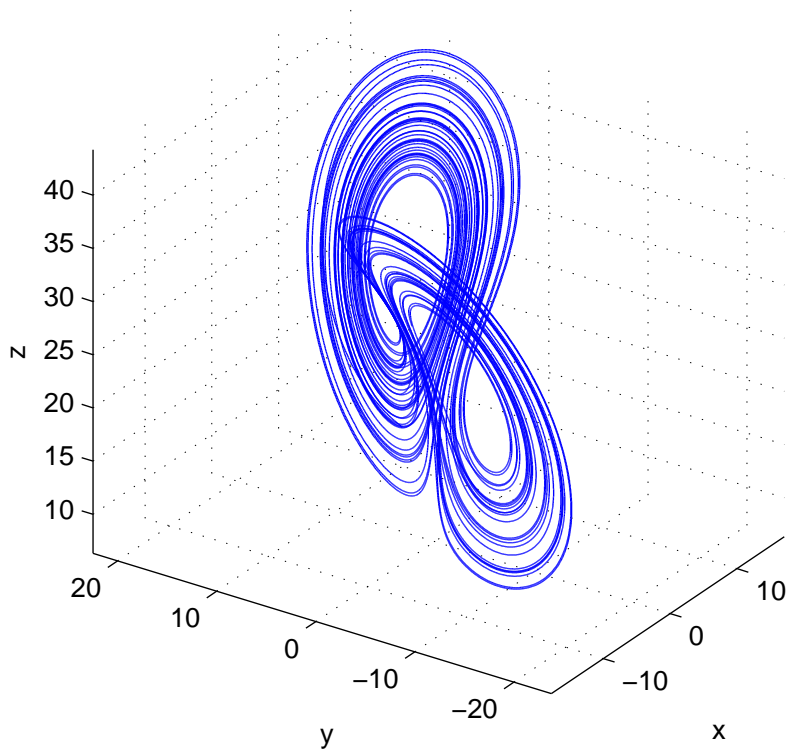}
\caption{The Lorenz strange attractor feeding the gauged pendulum.}
\label{fig:2}
\end{figure}
\end{center}

For certain parameter values and initial conditions, the Lorenz
system is known to exhibit chaos. For instance, choosing the
parameter values $\sigma=10,\,r=28,\,b=8/3$, the initial conditions
$x(0)=10,\,y(0)=10,\,z(0)=10$, and simulating for $t_f=50$ time
units, yields the trajectory depicted by Fig.~\ref{fig:2}.

\begin{center}
\begin{figure}[h]
\includegraphics[width=6in]{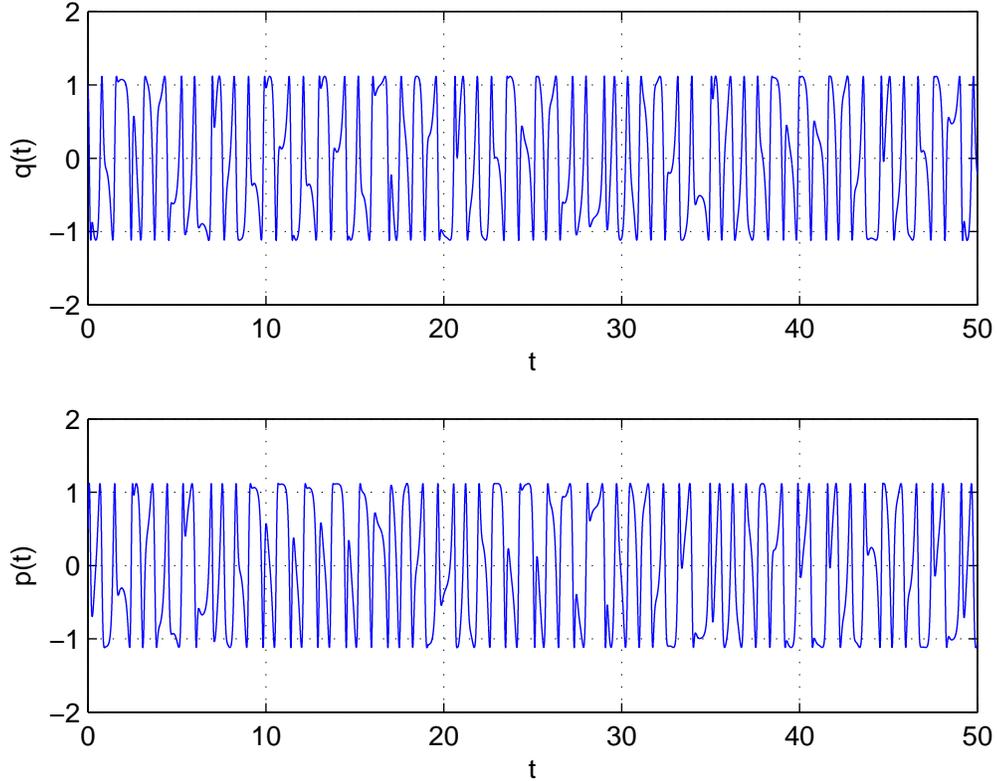}
\caption{A seemingly irregular behavior of a gauged pendulum fed by
a chaotic process.} \label{fig:3}
\end{figure}
\end{center}

Let us now examine the time history of $p$ and $q$, shown in
Fig.~\ref{fig:3}, and ask: Do $q$ and $p$ exhibit chaotic behavior?
To answer this seemingly trivial question (without using a
comprehensive mapping of the phase space using Poincar\`{e}
sections) , we shall resort to the common ``engineering"
interpretation of chaos, although more mathematically-rigorous
definitions, related to the destruction of KAM tori \cite{aulbach}
or the Kolmogorov-Sinai entropy \cite{frigg}, do exist. As Strogatz
says in reference \cite{strogatz}, ``no definition of the term chaos
is universally accepted yet, but almost everyone would agree on the
three ingredients used in the following working definition''. These
three ingredients are:

\begin{enumerate}

\item\label{1} Aperiodicity: Chaos is aperiodic long-term behavior in a deterministic system. Aperiodic long-term behavior means that there are trajectories
which do not settle down to fixed points, periodic orbits, or
quasiperiodic orbits as $t \to \infty$. \footnote{For the purposes
of this definition, a trajectory which approaches a limit of
$\infty$ as $t \to \infty$ should be considered to have a fixed
point at $\infty$.}

\item \label{2} Sensitive dependence on initial conditions: Nearby
trajectories separate exponentially fast, i.e., the system has a
positive Lyapunov characteristic exponent (LCE).

Strogatz notes that he favors additional constraints on the
aperiodic long-term behavior, but leaves open what form they may
take. He suggests two alternatives to fulfill this:

\item \label{3} Requiring that there exists an open set of initial conditions having aperiodic trajectories, or
\item \label{4} If one picks a random initial condition $x(t_0)=x_0$ then there must be a nonzero chance of the associated trajectory $x(t,x_0)$ being aperiodic.
\end{enumerate}

Returning to Fig.~\ref{fig:3}, we see that items \ref{1}, \ref{3}
and \ref{4} in Strogatz's list are satisfied: $p$ and $q$ exhibit
aperiodic behavior, the open set of initial conditions guaranteeing
aperiodic trajectories for $\sigma=10,\,r=28,\,b=8/3$ are
$x_0,y_0,z_0,p_0,q_0\in\mathbb R \backslash \{0\}$, and hence for
randomly selected initial conditions $p$ and $q$ will be aperiodic.
The only remaining test is to calculate the LCEs, denoted by
$\lambda_i,\,i=1,\ldots,n$. However, as shall be illustrated
shortly, calculation of the LCEs may be problematic for system
(\ref{lorentz}).

\begin{center}
\begin{figure}[h]
\includegraphics[width=4.5in]{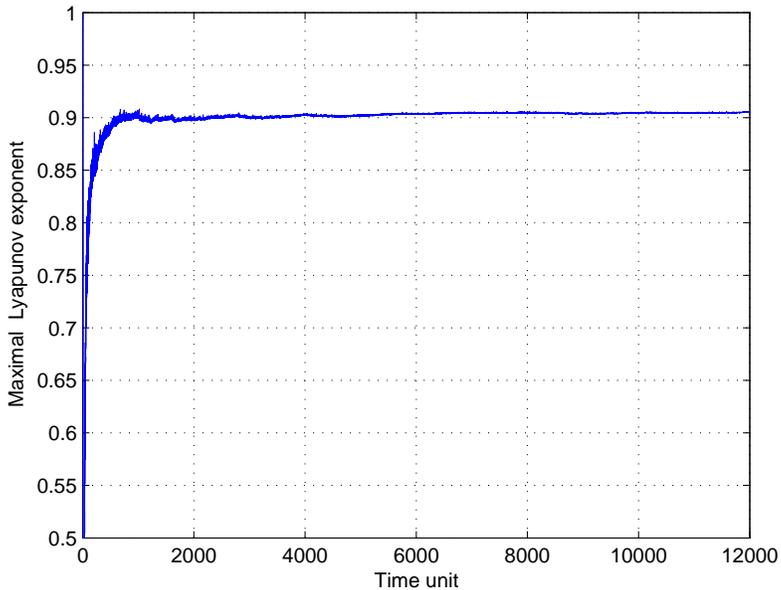}
\caption{The maximal Lyapunov characteristic exponent for a
Lorenz-fed gauged pendulum system.} \label{fig:maxlce}
\end{figure}
\end{center}

First, we should note that some authors endorse the calculation of
the \emph{maximal} Lyapunov exponent in order to establish the
presence of chaos. For example, Ref.~\cite{sandor} states that ``it
is well-known that the ordered or the chaotic property of an orbit
is characterized by the largest Lyapunov characteristic exponent".
This approach, however, is misleading for system (\ref{lorentz}). To
illustrate this fact, we have calculated the maximal Lyapunov
exponent for (\ref{lorentz}) using the standard method developed by
\cite{shimada} and \cite{wolf}. The result is depicted by
Fig.~\ref{fig:maxlce} for an integration period of $12,000$ time
units. It is seen that the maximal LCE satisfies
$\max{_i}\lambda_{i}\approx 0.9$, which is the well-known maximal
LCE of the Lorenz system. Hence, according to the rationale of
\cite{sandor}, system (\ref{lorentz}) is chaotic - or is it?

\begin{center}
\begin{figure}[h]
\includegraphics[width=4.5in]{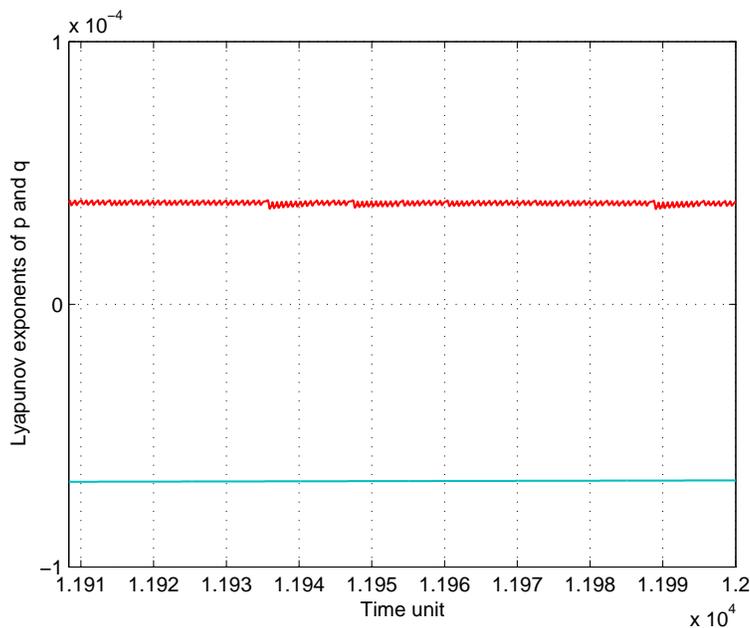}
\caption{The Lyapunov exponents contributed by the states $p$ and
$q$.} \label{fig:newlces}
\end{figure}
\end{center}

For a more rigorous analysis, the entire spectrum of LCEs should be
examined. Since the LCE spectrum of the Lorenz system is well-known
(the phase space contraction satisfies the relation $\sum_i
\lambda_i = \nabla\cdot [\dot x, \dot y, \dot z] = -(\sigma+b+1) =
-13.667$), let us concentrate on the additional LCEs contributed by
$p$ and $q$. A magnified view of these LCEs is shown in
Fig.~\ref{fig:newlces}. One of these LCEs is smaller than zero,
while the other one assumes the value of $~ 4\cdot 10^{-5}$, which
allegedly indicates that the additional states are also chaotic.

\begin{center}
\begin{figure}[h]
\includegraphics[width=6in]{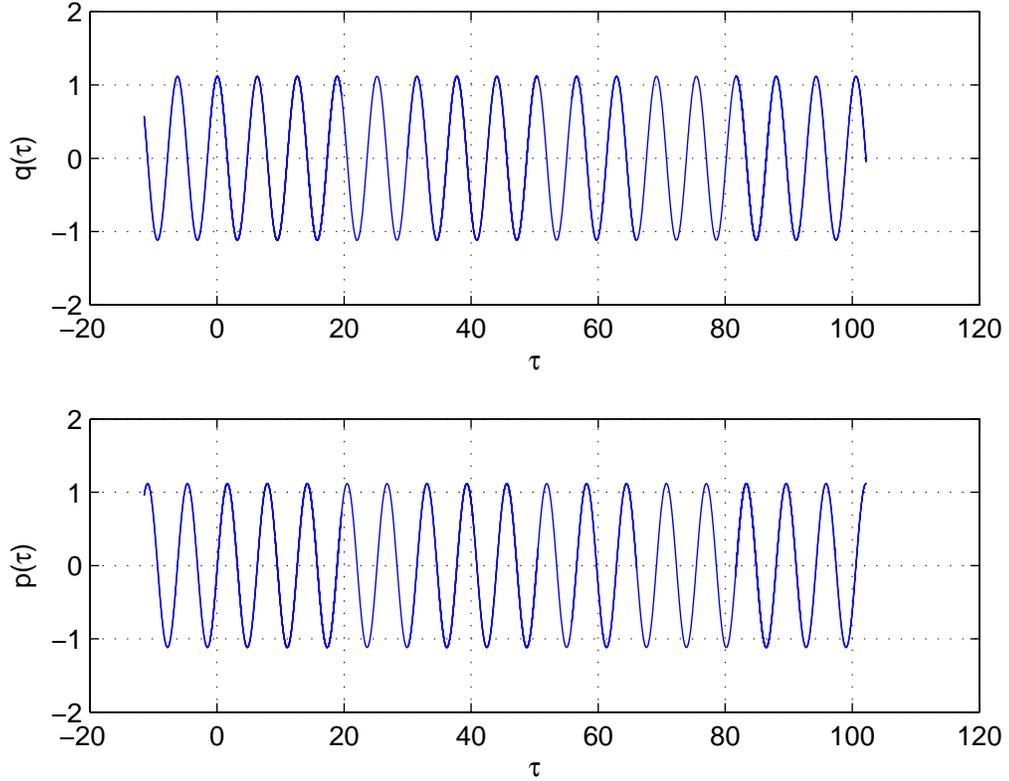}
\caption{The seemingly irregular behavior of the Lorenz--fed gauged
pendulum, shown in Fig.~\ref{fig:3}, can be regularized into
harmonic oscillations by a rescriptive gauge transformation.}
\label{fig:4}
\end{figure}
\end{center}

However, this is a mere illusion resulting from the fact that the
calculation process of the LCEs is affected by the truncation and
round-off errors of the numerical integration routine used to
simultaneously integrate the extended phase space of the original
and linearized systems.\footnote{This causes the Lyapunov exponents
themselves to exhibit a chaotic behavior; most high-order
integrators are chaotic maps, as pointed out in \cite{cartwright}.
This may be viewed a manifestation of the uncertainty principle.}
One may view this phenomenon as \emph{pseudochaos}
\cite{lowenstein}; the truth regarding ``chaos" in system
(\ref{lorentz}) can be plainly revealed by realizing that
(\ref{lorentz}) complies with the gauged-pendulum formalism, and can
hence be subjected to a rescriptive gauge transformation of the from
$d\tau  = y dt$. This transformation will transform
(\ref{lorentz1}), (\ref{lorentz2}) into $q^{\prime} = p,\,p^{\prime}
= -q$, which is an integrable system and hence \emph{cannot} exhibit
chaos. This observation is illustrated in Fig.~\ref{fig:4}, showing
plots of $q$ and $p$ as a function of $\tau$. Thus, in contrast to
the prediction of the common engineering interpreting of chaos and
the chaos detection tools thereof, the rescriptive gauge
transformation shows that the temporal behavior of signals cannot
always be used to predict the presence of chaos. This observation
calls into being the concept of \emph{partial chaos} \cite{campa},
meaning that in a given system, both chaotic and regular signals may
co-exist, even if the chaotic states overshadow the regular behavior
of the other states.

Another important conclusion concerns the system \emph{observables}.
Observables, or outputs, is a subset of state variables, $\vec
z,\,\dim\vec z = l \le \dim \vec x = n$, determined by the
\emph{output map}, $\vec{\mathfrak  O}: \mathbb R^n\to\mathbb R^l$,
such that $\vec z = \vec{\mathfrak O}(\vec x)$, and an
\emph{observation scale}, $\mathfrak T \in\mathbb R$, such that
$\vec z:\mathbb R\to \mathbb R^l$. If $\mathfrak T = t$, the
observation process is \emph{temporal} and the observable scale is
merely the time. Our simple example shows that temporal observations
may be misleading when used to detect chaos, even when using a
seemingly rigorous test such as the LCE spectrum. A fictitious
observer using $\mathfrak T=\tau$ as the scale would have not
suspected that the Lorenz-fed gauged pendulum is a chaotic process.

We further conclude that rescriptive gauge transformations may be
used to isolate self-similarities of a dynamical systems. In our
example, the Lorenz system remains scale-invariant; i.~e. its
Hausdorff dimension does not depend on the scale. However, the
Lorenz-fed pendulum is \emph{not} scale invariant, and hence is a
regular process in disguise.

\begin{example}[Stochastic signals, coding, and Kolmogorov
complexity]\end{example}

The preceding example illustrated the fact that the gauged pendulum
concept may be used to order pseudochaotic behavior. This is, in
fact, only an understatement of the potential of rescriptive gauge
theory; this theory can be used not only for ordering pseudochaotic
signals, but moreover, transform seemingly stochastic signals into
deterministic ones.

Our final example is therefore concerned with illustrating how
rescriptive gauge symmetry, and in particular a simple gauged
pendulum, may be used to establish some key ideas in modern
information and coding theory through the well-known notion of
Kolmogorov complexity.

The Kolmogorov complexity (also known as Kolmogorov-Chaitin
complexity, stochastic complexity, and algorithmic entropy) of an
object is a measure of the computational resources needed to specify
the object \cite{kunin3,li,chaitin}. In other words, the complexity
of a string is the length of the string's shortest description in
some fixed description language. It can be shown that the Kolmogorov
complexity of any string cannot be too much larger than the length
of the string itself. Strings whose Kolmogorov complexity is small
relative to the string's size are not considered to be complex. The
sensitivity of complexity relative to the choice of description
``language" is what the current example is about. To that end,
consider the gauge pendulum
\begin{equation}\label{stoch}
    \dot q = w p,\quad \dot p  = -w q
\end{equation}
where here the rescriptor $w$ is a \emph{band-limited white noise},
that is, a white noise going through a zero-order hold with some
sampling frequency $T_w$ and power spectral density $W$. Model
(\ref{stoch}) can be de-rescribed by $d\tau = w dt$.

\begin{center}
\begin{figure}[h]
\includegraphics[width=6in]{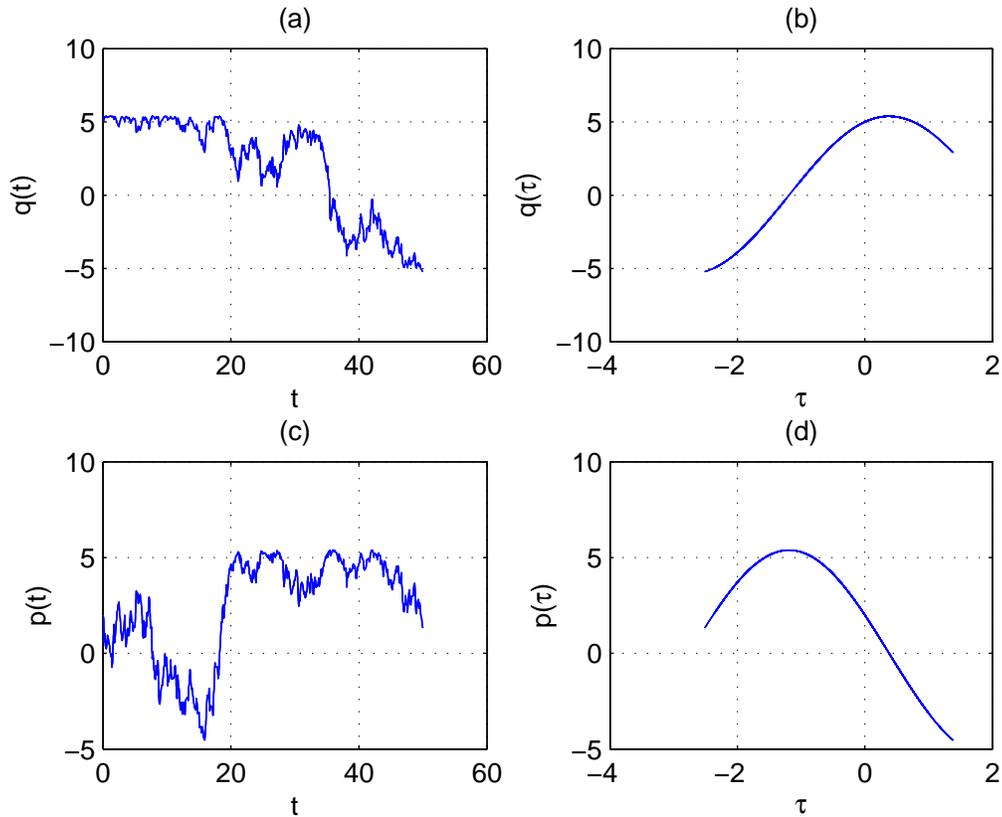}
\caption{``Stochastic" signals transformed into harmonic
oscillations by a rescriptive gauge transformation.} \label{fig:6a}
\end{figure}
\end{center}

Let us compare the representation of the ``strings" $q$ and $p$
using the ``languages" $t$, time, and $\tau$, a random walk obtained
by integrating $w$ (i.~e., a stochastic signal in its own right).
This comparison is depicted in Fig.~\ref{fig:6a} for $T_w = 0.1$
time units and $W = 0.1$. Fig.~\ref{fig:6a}a shows the signal
$q(t)$, which should be compared to the signal $q(\tau)$, shown in
Fig.~\ref{fig:6a}b. Similarly, compare $p(t)$, Fig.~\ref{fig:6a}c,
to $p(\tau)$, Fig.~\ref{fig:6a}d.

Although $q(t)$ and $p(t)$ seem stochastic and therefore
Kolmogorov-complex in the ``language" $t$, their alleged complexity
vanishes when the ``language" $\tau$ is used, and the stormy
stochasticity vanishes into harmonic oscillations, implying
much reduced Kolmogorov complexity. 
This phenomenon has
practical value in terms of coding theory: Signals may be coded
using the ``code" $t$ and de-coded using the ``key" $\tau$.


\section{Descriptive Gauge Symmetry}

A more ``benign" gauge symmetry can be detected by applying the
rationale of the Yang-Mills gauge theory of infinite-dimensional
systems on finite-dimensional dynamics. In this case, the problem
can be defined as follows: Given a rescribed dynamical system, of
the form (\ref{resc1}),

\[
    \mathfrak F_t\circ \vec f = \vec g(\vec x,\vec u(\vec x,t)
    )=\frac{d\vec
    x}{dt},
\]
find a \emph{descriptive gauge function} $\Psi_j$, possibly
different for each rescriptor component $u_i$,  satisfying
\begin{equation}\label{psid1}
    d\Psi_j= H_i(\vec x, d \vec x, \dot{\vec x},d\dot{\vec x},u_i(\vec x,t),t, dt),
\end{equation}
such that the \emph{gauge automorphism}
\begin{equation}\label{desc2}
    \mathfrak F_{\Psi}\circ \varphi(\xi_i(\vec x_0,t_0),t) = \varphi(\xi_i(\vec x_0,t_0),t)
\end{equation}
holds for some $\xi_i(\vec x, t),\,\in\mathcal
F,\,i\in[1,\ldots,n]$, where $\mathcal F$ is some abstract
configuration manifold embedded in $\mathbb R^n$.
If $\exists\Psi_j,\,j\in[1,\ldots,n]$ satisfying (\ref{psid1}) such
that (\ref{desc2}) holds, then we shall say that system
(\ref{resc1}) exhibits \emph{descriptive gauge symmetry} under the
\emph{descriptive gauge transformation} (\ref{psid1}). In this case
each $\Psi_j$ - and not the rescriptor, as in rescriptive gauge
symmetry - becomes either a static or a dynamic \emph{descriptive
gauge function}.

A \emph{descriptive gauge symmetry of order $k$} or simply
\emph{partial descriptive gauge symmetry} comes about when the
descriptive gauge transformation does not affect $k$ state
variables, $k< n$, viz.
\begin{equation}\label{resc2}
    \mathfrak F_{\Psi}\circ \varphi(\xi_i(\vec x,t_0),t) = \varphi(\xi_i(\vec x,t_0),t)
,\quad i\in \mathbb N^p.
\end{equation}
\subsection{Newtonian Systems Revisited}
\label{sec:newton2}
 Consider the Newtonian system
\begin{equation}\label{sys}
    \ddot {\vec q}(t)  + \vec f[\vec q(t)] = \vec u[\vec q(t),\dot{\vec
    q}(t),t]
\end{equation}
with the configuration manifold $Q$, $\vec q \in Q$, and the tangent
bundle $TQ = \mathbb R^{n/2}\times\mathbb R^{n/2}$, so that $(\vec
{q},\dot{\vec q})\in TQ$, $\vec f:Q\rightarrow \mathbb R^{n/2}$, and
$\vec u:Q\times \mathbb R\rightarrow \mathbb R^{n/2}$ is the
rescriptor. Let
\begin{equation}\label{gamma}
    \vec q = \vec \gamma[\vec x (t),t]
\end{equation}
be the solution of (\ref{sys}), where $\vec x:\mathbb R \rightarrow
\mathcal M \subseteq\mathbb R^n$ are the variational coordinates.
The velocity vector field is then given by the Lagrangian
derivative\footnote{A derivative taken with respect to a moving
coordinate system. Alternatively, this operation is sometime
referred to as the substantive derivative or Stokes derivative.
Fluid dynamicists prefer the notation $d/dt = \partial/\partial
t+\vec v \cdot\nabla$, where $\vec v$ is the velocity vector field.}
\begin{equation}\label{beta}
   \dot{ \vec q} = \vec \beta [\vec x (t),\dot{\vec x}(t), t]=\frac{\partial \vec \gamma[\vec x (t),t]}{\partial t}+\frac{\partial\vec \gamma[\vec x (t),t]}{\partial
   \vec x}\dot{\vec x},
\end{equation}
or, stated in terms of field theory, the \emph{gauge covariant
derivative} of the configuration vector field,
\begin{equation}\label{}
    \D_t\vec \gamma = \partial_t \vec \gamma + \grad_{\vec x}\vec \gamma\cdot  \D_t\vec
    x,
\end{equation}
and the velocity vector field,
\begin{equation}\label{}
    \D_t\vec \beta = \partial_t \vec \beta + \grad_{\vec x}\vec \beta\cdot  \D_t\vec
    x.
\end{equation}
 Denote the convective (sometimes also called
advective) term by
\begin{equation}\label{gauge}
    \vec\Psi \doteq \frac{\partial\vec \gamma[\vec x (t),t]}{\partial
   \vec x}\dot{\vec x}.
\end{equation}
Substituting (\ref{gamma}) and (\ref{beta}) into (\ref{sys})
transforms (\ref{sys}) into the Gauss-Poisson equations
%
  \begin{equation}\label{gvegauge2}
    \frac{ d{\vec x}(t)}{dt}=P^T(\vec x)\left\{\left[\frac{\partial\vec \gamma}{\partial\vec x}\right]^T\left(\vec u-\frac{d\vec \Psi}{dt}\right)-\left[\frac{\partial\vec{\beta}}{\partial\vec x}\right]^T\vec\Psi\right\}
\end{equation}
where $P$ is the $n\times n$ skew-symmetric Poisson matrix, whose
entries are the Poisson brackets, $\{x_i,x_j\}$.

We note that $\vec q$ -- the physical trajectory on the
configuration manifold -- remains invariant under any selection of
$\vec \Psi$. Thus, we are in the liberty of choosing a descriptive
gauge function vector of the form
\begin{equation}\label{psid}
    \vec \Psi = \left\{\begin{array}{l}
                         \vec W(\vec x,\vec u(\vec x,t), \dot{ \vec
    x},t) ,\: \vec u(\vec x,t)\ne 0 \\
                         \vec 0  ,\qquad \qquad \qquad\,\, \,\,\,\,\,\, \vec u(\vec x,t)=0,\\
                       \end{array}\right.
\end{equation}
inducing an affine descriptive gauge transformation of the form
%
%
\begin{equation}\label{}
d\vec\Psi   =  \vec H(\vec x, d \vec x,\dot{\vec x}, d \dot{\vec
x},t)dt,
\end{equation}
with
\begin{equation}\label{}
    H(\vec x, d \vec x,\dot{\vec x}, d \dot{\vec
x},t) \doteq \partial_t \vec \Psi +\grad_{\vec x}\vec \Psi \cdot
{\vec x}+\grad_{\dot{\vec x}}\vec \Psi\cdot \dot{\vec x},
\end{equation}
 such that the gauge automorphism
\begin{equation}\label{desc1}
    \mathfrak F_{\Psi}\circ \varphi(\vec q(\vec x_0,t_0),t) = \varphi(\vec q(\vec x_0,t_0),t)
\end{equation}
holds. We note that the gauge $\vec\Psi$ in the Newtonian context
has dimensions of velocity, and hence can be referred to as the
\emph{gauge velocity}. Recall that we have made a similar
observation regarding rescriptive gauge symmetry in Newtonian
systems (cf. \S\ref{sec:newtonrescriptive}).

We see that Newtonian systems exhibit partial descriptive gauge
symmetry, so that trajectories in the configuration space remain
\emph{invariant} under a selection of a particular descriptive gauge
function. Stated more eloquently, $    \vec q = \vec \gamma[\vec x
(t),t] $ remains invariant under the symmetry transformation
%
%
\begin{equation}\label{rule}
    {\partial_t \vec \gamma} \mapsto  {\D_t \vec \gamma}=  \partial_t \vec \gamma +
    \vec\Psi.
\end{equation}
The gauge group $\mathcal G$ therefore consists of real valued
functions on $\mathbb R^{n/2}$, with the group operation being
addition. An element $\vec \Psi$ acts on the velocity vector field
according to the rule (\ref{rule}).

 Eq. (\ref{gvegauge2}) is not necessarily
integrable, and may posses no ``classical" integrals whatsoever.
However, regardless of the particular properties of the original
system (\ref{sys}), the variational system (\ref{gvegauge2}) must
\emph{always} satisfy the constraint (\ref{gauge}). If we choose
$\vec \Psi \equiv \vec 0$, then (\ref{gauge}) becomes
\begin{equation}\label{gauge}
    \frac{\partial\vec \gamma[\vec x (t),t]}{\partial
   \vec x}\dot{\vec x} = \vec 0,
\end{equation}
which may be viewed as a \emph{hidden integral} emanating from the
descriptive gauge symmetry.

We emphasize that our hidden symmetry is not confined to systems in
which the homogenous solution of $\ddot {\vec q} + \vec f(\vec q)=0$
can be found. Although there are important realms of science in
which a solution to this system does exist - the most notable being
the case where $\vec f = \nabla R$, where $R$ is an inverse square
gravitational potential emerging in Keplerian orbital mechanics
\cite{efroimsky2} - in many other instances $\vec \gamma$ cannot be
found in closed form. This stems from the fact that the distinction
between $\vec f$ and $\vec u$, the rescriptor, is really an
artificial one; we can always take $\vec \chi = \vec u - \vec f$, so
that now
\begin{equation}\label{sys2}
    \ddot {\vec q}(t)    = \vec \chi[\vec q(t),\dot{\vec q}(t),t ].
\end{equation}
Letting $\vec x = [\vec x_1^T \quad \vec x_2^T]^T$ entails
\begin{equation}\label{gamma2}
     \vec \gamma[\vec x (t),t] = \vec x_1 t+\vec x_2.
\end{equation}
%
In this case the Gauss-Poisson equations are simply
\begin{equation}\label{}
    \dot{\vec x} = \left[%
\begin{array}{c}
  \dot{\vec x}_1 \\
  \dot{\vec x}_2 \\
\end{array}%
\right]= \left[%
\begin{array}{cc}
  0_{n/2\times n/2} & I \\
  I & -tI \\
\end{array}%
\right]\left[%
\begin{array}{c}
  \vec \Psi \\
  \vec \chi (\vec x_1, \vec x_2, \vec \Psi ,t)-\dot{\vec\Psi} \\
\end{array}%
\right],
\end{equation}
where $I$ is an $n/2\times n/2$ identity matrix and $0_{n/2\times
n/2}$ is an $n/2\times n/2$ zero matrix.

For $\vec\Psi \equiv 0$, our hidden integral re-emerges, assuming
the particulary simple form
\begin{equation}\label{const}
    \dot{\vec x}_1 t+\dot{\vec x_2} = \vec 0,
\end{equation}
for which
\begin{subequations}
 \label{simple}
\begin{eqnarray}
  \dot {\vec x}_1 &=& \vec \chi (\vec x_1, \vec x_2 ,t)\\
  \dot {\vec x}_2 &=& -t  \vec \chi(\vec x_1, \vec x_2, t).
\end{eqnarray}
\end{subequations}
Eqs.~(\ref{simple}) are particularly amenable for numerical
integration, because in this form constraint (\ref{const}) should be
 satisfied; however, due to numerical round off errors, this
 constraint is violated. An improved numerical integration
 may be achieved if the integration scheme itself is
 forced to satisfy this constraint during the integration process.

 We shall illustrate these observations in \S\ref{sec:examplesdes},
 discussing a few numerical examples showing how descriptive
 gauge symmetry may be used to improve numerical integration of
 ordinary differential equations. Our last section before dwelling
 upon actual examples deals with reduction in
 descriptive gauge theory.

\subsection{Reduction using Descriptive Gauge Symmetry}

Equivalently to \S\ref{sec:berkovich}, we shall conceive a process
for reduction using descriptive gauge symmetry. To that end, we
re-write Eq.~(\ref{gvegauge2}) into the following form:
\begin{equation}
  \dot {x}_j(t) = \sum_{i=1}^{n/2} f_{ji}(x)(u_i-\dot\psi_i)-
  \sum_{i=1}^{n/2}
  g_{ji}(x)\psi_i,\quad j = 1\ldots n
\end{equation}
This yields $n/2$ integrals $x_k,\,k\in \mathbb N^{n/2}$ (there are
$n!/(n/2!)^2)$ possible combinations of constants of motion obtained
by concomitant $n!/(n/2!)^2$ descriptive gauge function components)
obtained by solving the $n/2$ first-order ODEs
\begin{equation}
  \sum_{i=1}^{n/2} f_{ji}(x)\dot\psi_i+
  \sum_{i=1}^{n/2}
  g_{ji}(x)\psi_i =   \sum_{i=1}^{n/2} f_{ji}(x)u_i.
\end{equation}
The freedom to reduce system (\ref{gvegauge2}) stems from the
existence of the descriptive gauge function $\vec\Psi$. If we fix
the gauge - a straightforward selection would be $\vec \Psi = \vec
0$, as Lagrange himself had advocated in his memoirs
\cite{lagrange1,lagrange2,lagrange3} - this freedom will be lost,
and hence the possibility for reduction. This process is illustrated
in the following section.
\newpage
\subsection{Illustrative Examples}

\label{sec:examplesdes}

\begin{example}[Reduction using descriptive gauge
symmetry]\end{example}

Our first example is a simple one, illustrating the concept of
reduction using descriptive gauge symmetry. To that end, consider
the one-dimensional, second-order ODE
\begin{equation}
      \label{ }
          \ddot q(t)  + q(t)   =   \sin(t),\quad q(t_{0})=q_{0},\quad
      \dot{q}(t_{0})=\dot{q}_{0}.
\end{equation}
The general solution is
\begin{equation}
      \label{general_sol}
      q(t)   =   x_{1}(t)  q_{1}(t)+x_{2}(t)  q_{2}(t),
\end{equation}
where $q_{1}(t)$ and $q_{2}(t)$ are the fundamental solutions
    \begin{equation}\label{q1q2}
      q_{1}(t) =  \cos(t),\quad      q_{2}(t) =  \sin(t).
    \end{equation}
Taking $\Psi = \Psi (t)$ to be a time-dependant descriptive gauge
function, the Gauss-Poisson equations (\ref{gvegauge2}) assume the
simple form
\begin{subequations}\label{x1x2}
      \begin{equation}\label{x1x21}
      \dot{x}_{1}(t)  =   {\Psi(t)\dot{q}_{2}(t)-q_{2}(t)
              [ u(t)-\dot{\Psi}(t)
               ]         }
      \end{equation}
      \begin{equation}\label{x1x22}
      \dot{x}_{2}(t)  =   {q_{1}(t)
                   [ u(t)-\dot{\Psi}(t)
                    ]-\Psi(t)\dot{q}_{1}(t) }
      \end{equation}
\end{subequations}
Substituting (\ref{q1q2}) into (\ref{x1x2}) yields ODEs for two
possible descriptive gauge functions that will transform system
(\ref{x1x2}) into a single ODE. Thus, letting $k_1$ and $k_2$ denote
arbitrary integration constants, taking $\Psi(t) =
\Psi_1(t)=[t/2-\sin(2t)/4+k_1]/\sin t $ will reduce  (\ref{x1x2})
into
\bse \label{one}\begin{eqnarray}
   x_1 &=& const. \\
  \dot x_2 &=& -1/2 [\cos t \sin t -t-2 k_1]/\sin^2 t\label{onetwo}
\end{eqnarray}  \ese
and taking $\Psi(t) = \Psi_2(t) =[-\cos(2 t)/4+k_2]/\cos t $ will
reduce (\ref{x1x2}) into
\bse \label{two}\begin{eqnarray}
    x_2 &=& const. \\
  \dot x_1 &=& -1/4 [2\cos ^2 t -1-4 k_2)/\cos^2 t \label{twoone}.
\end{eqnarray}  \ese
Both (\ref{onetwo}) and (\ref{twoone}) are readily solved by
quadrature. Systems (\ref{one}) and (\ref{two}) will of course both
yield the \emph{same} general solution (\ref{general_sol}); this is
what descriptive gauge symmetry is all about.

\begin{example}[Gauss-Poisson variables reduce Hamiltonian
drift]\end{example}
Consider the Hamiltonian
\begin{equation}\label{}
    \mathcal H = \frac{1}{2}(\dot q^2+q^2)
\end{equation}
of the harmonic oscillator
\begin{equation}\label{sho}
    \ddot q(t)  + q(t) = 0.
\end{equation}
We shall compare the numerical integration of this equation in two
cases. In the first case, the state variables are chosen in standard
form: $q_1 = q$, $q_2 = \dot q$, so that (\ref{sho}) becomes
\begin{equation}\label{mod1}
    \dot q_1 = q_2,\quad \dot q_2 = -q_1,
\end{equation}
while in the second case, (\ref{sho}) is re-written in the form
$\ddot q = -q$, and the
 solution is taken as $\gamma = x_1(t) t+ x_2(t)$, with gauge $\Psi = 0$.
 Per (\ref{simple}), this yields the state-space representation
 \begin{equation}\label{mod2}
    \dot x_1 = -x_1 t-x_2,\quad \dot x_2 = x_1 t^2+x_2 t.
\end{equation}
We used MATLAB's ODE45 integration routine, a 5th-order Runge-Kutta
integrator with an adaptive time step, to integrate (\ref{mod1}) and
(\ref{mod2}) with an integration tolerance of $10^{-5}$ for 5000
time units given $\mathcal H = 2.5$.

\begin{center}
\begin{figure}[h]
\includegraphics[width=6in]{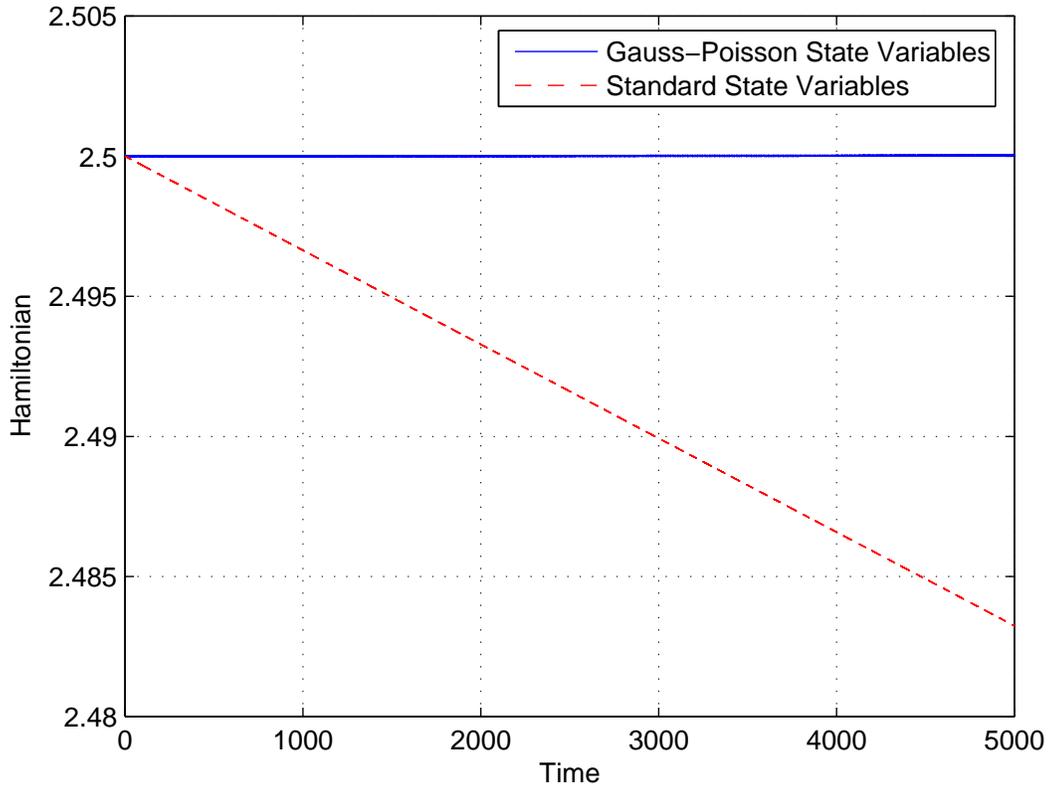}
\caption{Symplectifying a non-symplectic integrator using
Gauss-Poisson state variables and a zero descriptive gauge
function.} \label{fig:8}
\end{figure}
\end{center}

The time history of the Hamiltonian for both cases is plotted in
Fig.~\ref{fig:8}. Since the ODE45 routine is not a symplectic
integrator, the standard selection of states, Eq.~(\ref{mod1}),
causes the Hamiltonian to decrease with time at a rate of about
$5\cdot 10^{-4}$ units per time unit. This introduces artificial
numerical damping into the system, which causes the harmonic
oscillations to slowly damp out. However, the same system integrated
in form (\ref{mod2}) keeps the Hamiltonian fixed. This implies that
the Gauss-Poisson formalism may be used to ``symplectify"
non-symplectic integrators by a judicious selection of the
descriptive gauge function.

\begin{example}[Descriptive gauge reduces integration errors]\end{example}

Our final example shows how descriptive gauge symmetry may me used
to reduce the numerical truncation error of ODE integration. We
shall ultimately show that using the Gauss-Poisson state variables
with an appropriate descriptive gauge can dramatically reduce the
numerical truncation errors, and show how such gauge can be found.

Consider, for example, the one-dimensional, second-order ODE
\begin{eqnarray}
      \label{eq:general_main}
      \ddot{q}(t)+2\xi\omega_{n}\dot{q}(t)+\omega_{n}^{2}q(t)  =  u(t),\quad
      q(t_{0})=q_{0},\;
      \dot{q}(t_{0})=\dot{q}_{0},
\end{eqnarray}
where $\omega_{n}$ is the natural frequency, $\xi$ is the damping
coefficient and the rescriptor $u(t)$ is piecewise continuous.
Assuming an underdamped case ($\xi<1$),
\begin{eqnarray}
      \label{eq:general_sol}
      q(t)   =   x_{1}(t)  q_{1}(t)+x_{2}(t)  q_{2}(t),
\end{eqnarray}
where $q_{1}(t)$ and $q_{2}(t)$ are the fundamental solutions
\begin{subequations}\label{eq:hom_solutions}
    \begin{equation}\label{eq:hom_solution1}
      q_{1}(t) =  e^{-\xi\omega_{n}t}\cos(\omega_{d}t)
    \end{equation}
    \begin{equation}\label{eq:hom_solution2}
      q_{2}(t) =  e^{-\xi\omega_{n}t}\sin(\omega_{d}t)
    \end{equation}
\end{subequations}
and $\omega_{d}=\omega_{n}\sqrt{1-\xi^{2}}$.
Taking $\Psi = \Psi (t)$ to be a time-dependant descriptive gauge
function, the Gauss-Poisson equations (\ref{gvegauge2}) assume the
simple form
\begin{subequations}\label{eq:general_div_eq}
      \begin{equation}\label{eq:general_div_eq1}
      \dot{x}_{1}(t)  =  \frac{\Psi(t)\dot{q}_{2}(t)-q_{2}(t)
              \left [ \begin{array}{c} u(t)-\dot{\Psi}(t)-2\xi\omega_{n} \Psi(t)
               \end{array} \right ]         }{w[q_{1}(t),q_{2}(t)]}
      \end{equation}
      \begin{equation}\label{eq:general_div_eq2}
      \dot{x}_{2}(t)  =  \frac{q_{1}(t)
                   \left [ \begin{array}{c} u(t)-\dot{\Psi}(t)-2\xi\omega_{n} \Psi(t)
                    \end{array} \right ]
                        -\Psi(t)\dot{q}_{1}(t) }{w[q_{1}(t),q_{2}(t)]}
      \end{equation}
\end{subequations}
where $w[q_{1}(t),q_{2}(t)]$ is the Wronskian determinant,
\begin{eqnarray}
      w[q_{1}(t),q_{2}(t)]= \left | \begin{array}{cc}
              q_{1}(t)       & q_{2}(t) \\
              \dot{q}_{1}(t) & \dot{q}_{2}(t)
      \end{array} \right |.
\end{eqnarray}
The initial conditions for this system are then
\begin{eqnarray}
      \label{eq:general_cond_on_c}
      \nonumber
      x_{1}(t_{0}) & = & \frac{-q_{0}\dot{q}_{2}(t_{0})+q_{2}(t_{0})\dot{q}_{0}-q_{2}(t_{0})\Psi(t_{0})}
                          {\dot{q}_{1}(t_{0})q_{2}(t_{0})-\dot{q}_{2}(t_{0})q_{1}(t_{0})} \\
      x_{2}(t_{0}) & = & -\frac{-q_{0}\dot{q}_{1}(t_{0})+q_{1}(t_{0})\dot{q}_{0}-q_{1}(t_{0})\Psi(t_{0})}
                          {\dot{q}_{1}(t_{0})q_{2}(t_{0})-\dot{q}_{2}(t_{0})q_{1}(t_{0})}
\end{eqnarray}
From the discussion in \S\ref{sec:newton2}, we know that there is
only a single solution for $q(t)$ for any given initial conditions;
thus, per the descriptive gauge symmetry, $q(t)$ must remain
invariant to any selection of the gauge function $\Psi$. However,
$\Psi$ may be used as a tuning function for mitigating the numerical
integration error. To that end, we define the numerical integration
error of some state variable $(\cdot)$ as the difference between the
true solution and the numerical solution:
\begin{eqnarray}
      e_{(\cdot)}=(\cdot)_{true}-(\cdot)_{numerical}
\end{eqnarray}
In this example we will demonstrate how to mitigate the numerical
integration error of a (fixed-step) 4th order Runge-Kutta integrator
(RK4) by several orders of magnitudes. This merit is achievable by
applying \emph{gauge-optimized integration}. To that end, let us
re-write Eqs.~(\ref{eq:general_div_eq}) into
\bse
      \label{eq:dummy1}
\begin{eqnarray}
      \dot{x}_{1}(t) & = & f(t,\Psi ) \\
      \dot{x}_{2}(t) & = & g(t,\Psi )
\end{eqnarray}
\ese
 Obviously, in the linear case discussed herein, the transformation
into the Gauss-Poisson equations has transomed the ODE integration
problem into a simple \emph{quadrature}, whose accuracy can be
controlled by a proper selection of a time-dependant descriptive
gauge function.

\begin{center}
\begin{figure}[h]
  \includegraphics[width=6.0in]{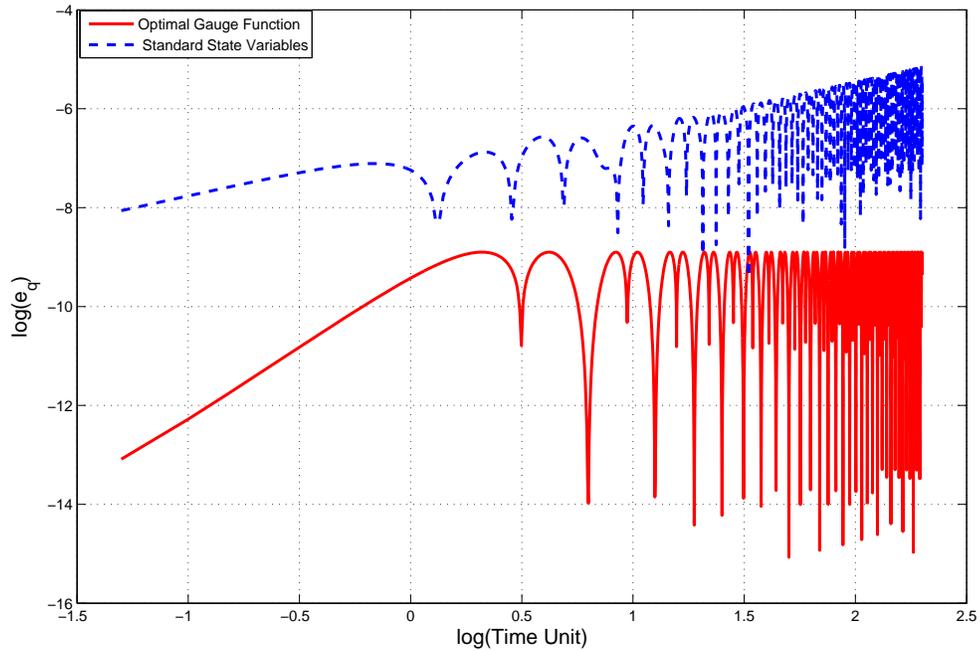}\\
  \caption{The numerical integration error of the gauge-optimized integration considerably reduces the integration error compared to a standard choice of state variables.}
  \label{fig:example1_1_grp}
\end{figure}
\end{center}

The integration errors resulting from numerically integrating
(\ref{eq:dummy1}) are given by \cite{stoer}
\bse
\begin{eqnarray}
      {e_{x_{1}}} & = &  -\frac{1}{90}h^{5}f^{(4)}[\xi,\Psi(\xi)]   \\
      {e_{x_{2}}} & = &  -\frac{1}{90}h^{5}g^{(4)}[\xi,\Psi(\xi)]
\end{eqnarray}
\ese
where $\xi\in(t_0,\,t)$.
The total integration error of $q(t)$ is now calculated as follows:
\begin{eqnarray}
      \label{eq:general_terror}
      \nonumber
      e_q & = & q_{1}(t){e_{x_{1}}}+q_{2}(t){e_{x_{2}}}\\
      & = &q_{1}(t) [-\frac{1}{90}h^{5}f^{(4)}(\xi,\Psi)] +q_{2}(t)
      [-\frac{1}{90}h^{5}g^{(4)}(\xi,\Psi)].
\end{eqnarray}
If some $\Psi^\star$ could be found for which $e_q\equiv 0,\ \forall
\xi\in(t_0,t)$, then the only remaining integration error of $q(t)$
would be the numerical round-off error. Indeed, such $\Psi^\star$
can be quite straightforwardly found. For example, if the forcing
term is of the form
\begin{eqnarray}
      \label{eq:genral_force}
      F(t)=\frac{a_{0}}{2}+\sum^{K_{c}}_{k=1}{a_{k}\cos(k\omega_{0}t)}
                            +\sum^{K_{s}}_{k=1}{b_{k}\sin(k\omega_{0}t)},
\end{eqnarray}
then $\Psi^\star$ may be chosen as a Fourier series as well:
\begin{eqnarray}
      \label{eq:phi_fourier}
      \Psi^\star (t)=\frac{A_{0}}{2}+\sum^{N_{c}}_{n=1}
             \left [ \begin{array}{c}
                   A_{n}\cos({n\omega_{0}t})
             \end{array} \right ]
             +\sum^{N_{s}}_{n=1}\left [ \begin{array}{c}
                   B_{n}\sin({n\omega_{0}t})
             \end{array} \right ]
\end{eqnarray}
Substituting (\ref{eq:phi_fourier}) into (\ref{eq:general_terror})
and solving for the coefficients by requiring $e_q=0$ yields, after
some algebra, that
\begin{eqnarray}
       \label{eq:phigen}
       \Psi^\star &=& \sum^{N}_{n=1}
            \frac{\ds \rho_{2}(n)}{\ds \rho(n)}\omega_{n}
               \left [  a_{n}\cos(nt\omega_{0})-b_{n}\sin(nt\omega_{0})
                 \right ]\nonumber \\ &+&
            \sum^{N}_{n=1} \frac{\ds \rho_{1}(n)}{\ds \rho(n)}\omega_{n}\xi
            \left [  a_{n}\sin(nt\omega_{0})+b_{n}\cos(nt\omega_{0})   \right ]
\end{eqnarray}
where
\begin{eqnarray}
      \nonumber
      \rho_{1} & = & 4[32\omega_{n}^6\xi^6+(-40\omega_{n}^6+32n^2\omega_{n}^4\omega_{0}^2)\xi^4
                          +(-24n^2\omega_{n}^4\omega_{0}^2+14\omega_{n}^6+10n^4\omega_{n}^2\omega_{0}^4)\xi^2 \\
      \nonumber
               & \ & \ \ +5\omega_{0}^6n^6+5n^4\omega_{n}^2\omega_{0}^4+7n^2\omega_{n}^4\omega_{0}^2-\omega_{n}^6] \\
      \nonumber
      \rho_{2} & = & -4n(16\omega_{n}^6\xi^6+(16n^2\omega_{n}^4\omega_{0}^2-16\omega_{n}^6)\xi^4+
                      (-12n^2\omega_{n}^4\omega_{0}^2+4\omega_{n}^6)\xi^2 \\
      \nonumber
               & \ & \ \ +5\omega_{0}^6n^6+11n^2\omega_{n}^4\omega_{0}^2+\omega_{n}^6+15n^4\omega_{n}^2\omega_{0}^4) \\
      \nonumber
      \rho & = &  (256\omega_{n}^8\xi^8+(320\omega_{n}^6\omega_{0}^2n^2-384\omega_{n}^8)\xi^6+(176\omega_{n}^8
             +160\omega_{n}^4\omega_{0}^4n^4-320\omega_{n}^6\omega_{0}^2n^2)\xi^4\\
      \nonumber
           & \ & \ \ +(-24\omega_{n}^8-120\omega_{n}^4\omega_{0}^4n^4+80\omega_{n}^6\omega_{0}^2n^2)\xi^2+20\omega_{n}^6\omega_{0}^2n^2
            +\omega_{n}^8+110\omega_{n}^4\omega_{0}^4n^4 \\
      \nonumber
           & \ & \ \
           +100\omega_{n}^2\omega_{0}^6n^6+25\omega_{0}^8n^8).
           \\
\end{eqnarray}
For illustration,  if we desire to integrate numerically
\begin{eqnarray}
      \ddot{q}+q=\sin 2t ,\quad q(0)=0,\quad \dot{q}(0)=0,
\end{eqnarray}
then the descriptive gauge function yielding minimum numerical
truncation will be, based on (\ref{eq:phigen}), simply
\begin{eqnarray}
     \label{eq:phi2}
     \Psi^{\star}=-\frac{40}{121}\cos(2t).
\end{eqnarray}
In Figure~\ref{fig:example1_1_grp} we depict a comparison of
integration errors between the gauge-optimized integration utilizing
the Gauss-Poisson equations with the optimal descriptive gauge
function (\ref{eq:phi2}) and the standard choice of state variables
$q_{1}=q, q_{2}=\dot{q}$. As can be plainly seen, the
gauge-optimized integration decreases the integration error by three
orders of magnitude in the examined time interval. Moreover, the
integration error using the standard state variables is diverging,
while the error of the gauge-optimized integration is bounded.
Therefore, for a larger time interval, the use of gauge-optimized
integration, utilizing the concept of descriptive gauge symmetry,
becomes increasingly important.

\section{Summary and Conclusions}

This paper described how gauge theory can be adapted for
finite-dimensional dynamical systems. We have defined gauge symmetry
in a very broad context, and distinguished between two fundamental
manifestations of gauge symmetry:

(i) Rescriptive gauge symmetry results from an action of a
one-parameter Lie group, yielding an Abelian Lie algebra. A
rescriptive gauge symmetry transformation is then an infinitesimal
change of the independent variable, which renders the system
integrable via reduction.

(ii) Descriptive gauge symmetry is an invariance of some
configuration space under a gauge transformation of the covariant
derivative. In this case the symmetry group consists of real-valued
functions on the Euclidean space, with the group operation being
addition.

The gauge conversation leads to a few practical conclusions. We
first note that gauge symmetry is ubiquitous in a myriad of
scientific fields. Gauge theory for finite-dimensional system may be
thus viewed as a generalization of dynamical systems theory into the
 realm of group theory, unifying various physical phenomenon
into simple generating models.

Furthermore, the gauge-theoretic tools may be used to improve our
understanding of chaos, randomness and their inter-relations. We
discussed a few simple examples showing how a change of scale can
lead to pattern evocation in seemingly chaotic and/or stochastic
systems.

Finally, gauge-theoretic tools are important for improving the
accuracy of numerical integration. The gauge freedom allows
re-shaping of the phase space so as to render it tractable for
numerical integration.

\section*{Acknowledgments}

I am in a debt of gratitude to Isaac Kunin and Michael Efroismky,
whose research inspired some paths of this endeavor. I am thankful
to Egemen Kolemen for pointing out manifestations of gauge symmetry
in finite-dimensional systems, and to Itzik Klein and Alex Kogan,
for making valuable suggestions.



\end{document}